\documentclass[11pt,a4paper]{article}
\usepackage{amsmath}
\usepackage[dvipdfmx]{graphicx}
\usepackage{xcolor}
\usepackage[caption=false]{subfig}
\usepackage{mathrsfs,mathtools}
\usepackage{physics,amssymb}
\usepackage{siunitx}
\usepackage{bm}
\usepackage{soul}
\usepackage{comment}
\usepackage{braket}
\usepackage{listings}

\usepackage{cases}
\usepackage{comment}
\usepackage{cancel}
\usepackage{cases}
\usepackage[utf8]{inputenc}
\usepackage{url}
\usepackage{float}
\usepackage{longtable}
\usepackage[normalem]{ulem}
\usepackage{xspace}
\usepackage{aas_macros}
\setstcolor{red}
\usepackage{jcappub}
\hypersetup{colorlinks=true
,urlcolor=DARKBLUE
,anchorcolor=DARKBLUE
,citecolor=DARKBLUE
,filecolor=DARKBLUE
,linkcolor=DARKBLUE
,menucolor=DARKBLUE
,linktocpage=true
,pdfproducer=medialab
}

\newcommand{\MS}{\mathrm{MS}}

\newcommand{\calO}{\mathcal{O}}

\newcommand{\bae}[1]{\begin{align} #1 \end{align}}

\newcommand{\beae}[1]{\begin{equation}\begin{aligned} #1 \end{aligned}\end{equation}}

\definecolor{MONZA}{HTML}{CF000F}
\definecolor{DARKBLUE}{HTML}{00008b}
\definecolor{DARKMAGENTA}{HTML}{8b008b}
\definecolor{DARKCYAN}{HTML}{008B8B}
\definecolor{DARKORANGE}{HTML}{FF8C00}

\begin{document}

\title{Primordial Black Hole formation from overlapping cosmological fluctuations}

\author[a]{Albert Escriv\`a}

\author[a]{and Chul-Moon Yoo}

\affiliation[a]{\mbox{Division of Particle and Astrophysical Science, Graduate School of Science,}  Nagoya University. Nagoya 464-8602, Japan}

\emailAdd{escriva.manas.albert.y0@a.mail.nagoya-u.ac.jp}
\emailAdd{yoo.chulmoon.k6@f.mail.nagoya-u.ac.jp}

\date{\today}

\abstract{We consider the formation of primordial black holes (PBHs), during the radiation-dominated Universe, generated from the collapse of super-horizon curvature fluctuations that are overlapped with others on larger scales. Using a set of different curvature profiles, we show that the threshold for PBH formation (defined as the critical peak of the compaction function) can be decreased by several percentages, thanks to the overlapping between two peaks in the profile of the compaction function. In the opposite case, when the fluctuations are sufficiently decoupled the threshold values behave as having the fluctuations isolated (isolated peaks). We find that the analytical estimates of Ref. \cite{universal1} can be used accurately when applied to the corresponding peak that is leading to the gravitational collapse. We also study in detail the dynamics and estimate the final PBH mass for different initial configurations, showing that the profile dependence has a significant effect on that.}

\maketitle
\flushbottom
%%%%%%%%%%%%%%%%%%%%%%%%%%%%%%%%%%%%%%%%%%%%%%%%%%%%%%%%%%%%%%%%%%%%%%%%%%%
\section{Introduction}

%%%%%%%%%%%%%%%%%%%%%%%%%%%%%%%%%%%%%%%%%%%%%%%%%%%%%%%%%%%%%%%%%%

A primordial black hole (PBH) is, supposed to be, a type of black hole that was not formed by the collapse of a sufficiently massive star, but created during the early Universe \cite{Carr:1974nx,Carr:1975qj} (see \cite{Escriva:2022duf,2022Galax..10..112Y} for recent reviews on the topic). PBHs, which are not made of baryonic matter \cite{Overduin:2004sz}, are good candidates for being the constituents of dark matter or a significant fraction of it \cite{Khlopov:2008qy,Frampton:2010sw,Bird:2016dcv,Carr:2009jm,Carr:2016drx,Carr:2020gox}. Moreover, they can be the source of gravitational waves emitted by binary black hole mergers \cite{LIGOScientific:2016aoc}.

There are several mechanisms that could have led to the production of PBHs (see \cite{Escriva:2022duf} for a detailed list), but the most considered one is from highly peaked density fluctuations in the post-inflationary early Universe \cite{Carr:1974nx}. If these fluctuations are sufficiently strong, the gravitational pull cannot be counteracted by the pressure force of the collapsing fluid and the expansion of the Universe, leading to the formation of black holes. This situation happens when the amplitude of the fluctuation is higher than a given threshold value.

If the in-homogeneities are generated by inflation and collapse during the radiation epoch, the statistics of PBHs are exponentially sensitive to their threshold of formation \cite{Carr:1975qj}. Clearly, this necessitates precision because of the exponential dependence involved. Numerical simulations are then needed since the threshold for PBH formation is indeed not a universal value, it depends on the equation of state but also the specific details of the shape of the curvature fluctuations and the scenario of formation considered \cite{Shibata:1999zs,Niemeyer2,IHawke_2002,Musco:2004ak,2014JCAP...01..037N,refrencia-extra-jaume,2018JCAP...08..041B,escriva_solo,universal1,Yoo:2020lmg,Escriva:2020tak,Kokubu:2018fxy,Musco:2018rwt,2021PhRvD.103f3538M,Escriva:2021pmf,Escriva:2022bwe,Franciolini:2022tfm,Harada:2022xjp,2022arXiv221115674E} (see \cite{2022Univ....8...66E} for a review). Although numerical simulations are needed to obtain the threshold with precision, some analytical estimates have been proposed \cite{Carr:1975qj,Harada:2013epa,universal1,Escriva:2020tak} \footnote{See also \cite{Papanikolaou:2022cvo,Papanikolaou:2023crz} based on \cite{Harada:2013epa} for a scenario with a time-dependent equation of state and in loop quantum cosmology respectively}, in particular some of them 
%that 
take into account the profile dependence and the equation of state \cite{universal1,Escriva:2020tak}, which have been accurately tested using the results from numerical simulations, in comparison with previous existing estimates. These analytical estimations were based on the use of the averaged critical compaction function, which was found to be a Universal value \cite{universal1} within a few percentages of deviation (being $2/5$ for the case of a radiation dominated Universe) for a perfect fluid with an equation of state given by $1/3 \leq w \leq 1$. In these estimations, the characteristic radius at which the compaction function $\mathcal C$ takes the maximal value plays a crucial role. 
The compaction function $\mathcal C$ is a function of the radial coordinate $r$ defined as the mass excess inside the sphere divided by the areal radius (see also \cite{Harada:2023ffo} for more details), and its usefulness was already noticed some time ago by Shibata \& Sasaki \cite{Shibata:1999zs} (confirmed later on with subsequent numerical studies \cite{refrencia-extra-jaume,Musco:2018rwt,universal1,Escriva:2020tak,2021PhRvD.103f3538M}).

In the literature, simulations have been mainly focused on studying initial conditions characterised by a single scale curvature fluctuations, that is, fluctuations that have been associated with a given amplitude and a specific length scale. 
In this case, the smallest radius $r$ at which the compaction function $\mathcal C(r)$ takes the maximal value is considered to be relevant for the PBH formation. 
However, the first peak of the compaction function $\mathcal C(r)$ as a function of the radial coordinate $r$ could be surrounded by other secondary peaks, and we may have to 
independently check the PBH formation criterion for each peak of $\mathcal C(r)$. 
Actually, the typical spherically symmetric profile of the curvature perturbation for the monochromatic power spectrum is known to be the sinc function, and 
the profile of the compaction function has infinitely repeated peaks in this case, as is explicitly shown in \cite{2020JCAP...05..022A}. 
Although the amplitudes for the secondary surrounding peaks are similar to that of the first peak, they will never lead PBH formation as is shown in \cite{2020JCAP...05..022A}.  
This fact shows that the naive criterion by using the value of the compaction function at the peak may not be appropriate for the case 
of multi-scale profiles. 
Then it would be necessary to carefully check the validity of existing PBH formation criteria for the multi-scale profiles.

\begin{figure}[h]
\centering
\includegraphics[width=3.0 in]{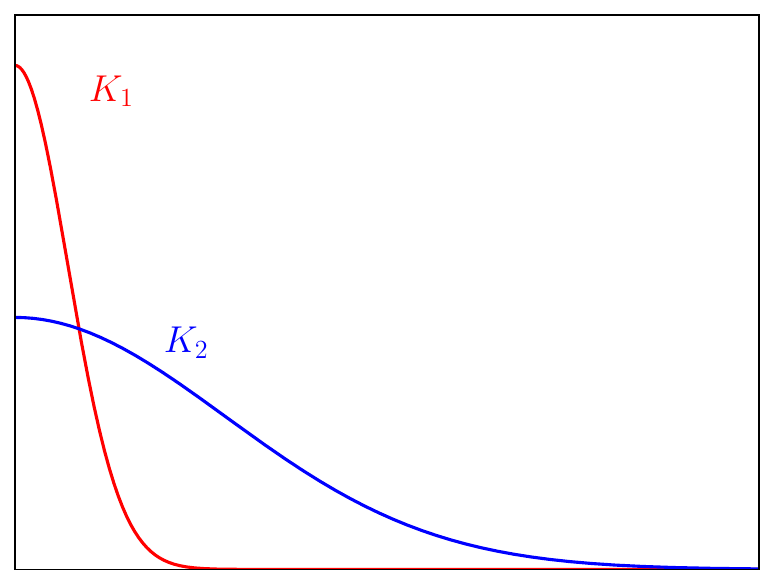}
\includegraphics[width=3.0 in]{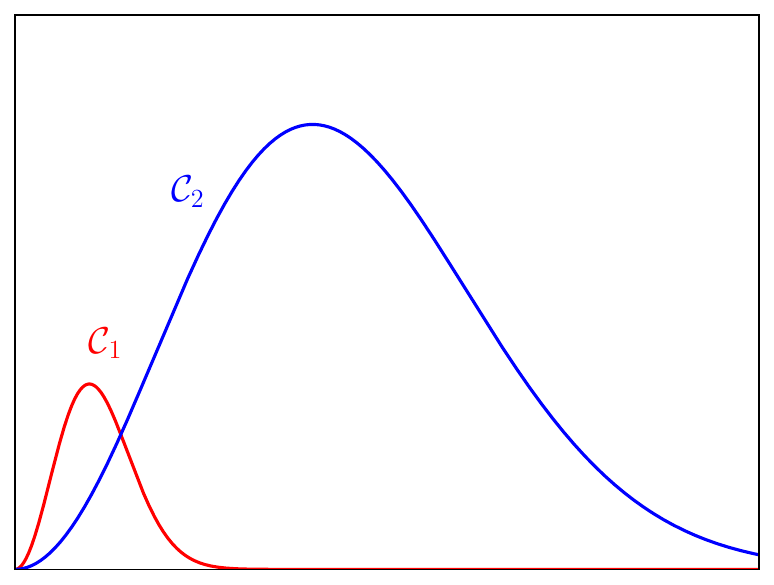}
\caption{Schematic picture of overlapping fluctuations in spherical symmetry (x-label radial coordinate). Left-panel: Both curvatures $K_{1}$ and $K_{2}$ are centrally peaked at the same location, but can have different strength and length-scale (in the figure, $K_2$ represents a curvature with a larger scale compared with the one of $K_1$). Right-panel: Schematic representation of the corresponding overlapped compaction function $\mathcal{C}_1$, $\mathcal{C}_2$ (mass excess) from $K_1, K_2$ (left-panel), which have different peak values and location of the maximum. Notice that, as numerical studies have shown, the existing criteria for PBH formation rely on the mass excess (compaction function $\mathcal{C}$) rather than the peaks of the curvature fluctuation $K$.}
\label{fig:overlaped_fluct}
\end{figure}

Throughout this paper, our analyses are based on the compaction function, which is roughly understood as the volume average of the density perturbation 
on the background spatially flat FLRW universe within a given radius (see \cite{refrencia-extra-jaume}). Therefore, assuming the two length scales of the fluctuations $s$ (short scale) and $l$ (large scale)
% fluctuations ($s$: short scale) and ($l$: large scale) 
are very separated from each other ($s \ll l$), the contribution from the short (long) scale fluctuation is trivially included in the compaction function evaluated in the long (short) scale. Then, naively, we expect that when $s \ll l$, both fluctuations are very decoupled, and PBH formation is mainly sourced by the one that is over-threshold in terms of the compaction function. On the other hand, when both fluctuations are at a similar scale $s \sim l$, the criterion of the PBH formation would be rather non-trivial.
%In this paper, we explore the scenario where the initial conditions for PBH formation are built by two curvature fluctuations, one superimposed to another one with a shorter length scale (in other words, overlapping fluctuations). \Blue{Naively, we expect that when two fluctuations ($s$: short scale) and ($l$: large scale) are very separated from each other ($s \gg l$ or $l \ll s$), both fluctuations are very decoupled, and PBH formation is mainly sourced by the one that is over-threshold. 
%On the other hand, when both fluctuations are at a similar scale $s \sim l$, the primordial black hole formation is enhanced due to the overlapping.} 
In Fig.~\ref{fig:overlaped_fluct} (left-panel), we show a schematic figure of overlapping compaction functions \footnote{It should be noted that, in our setting, the fluctuations in 
two different scales are always overlapping. However, the corresponding features in the compaction function do not necessarily significantly overlap with each other. Readers should clearly distinguish the overlapping of the fluctuations from the overlapping in the profile of the compaction function.}. %
Such a case was explored in~\cite{Nakama:2014fra} using relativistic numerical simulations for a specific curvature profile with fixed $l/s$ (for the case $l \gg s$) and peak amplitude, specifically focusing on the PBH dynamics denoted as ``double PBH formation" (see section $4.4$).
In this paper, we provide more comprehensive analyses as well as a systematic 
evaluation of the PBH formation criterion for overlapping curvature fluctuations using the compaction function. %\Blue{As we show in this work, overlapping curvature fluctuations can have associated two peaks in the compaction function as a function of the radial coordinate $r$ (see right-panel of Fig.~\ref{fig:overlaped_fluct}), which can be modulated with different heights and distances between them.}

%focusing on two parameters of the curvature profile $K$ to characterize the threshold of black hole formation. 

%But as we have mentioned, recent literature has found the compaction function $\mathcal{C}$ (mass excess) to be the useful quantity for characterizing the threshold of PBH formation. This is particularly relevant from the results of Ref. \cite{universal1}, where only one parameter was needed to infer if a fluctuation can lead to black hole formation. Therefore, in this work, we want to test such refined technique and procedure for this scenario of overlapped fluctuations (which can conveniently give us a compaction function with two peaks), considering general curvature profiles following the parametrization of \cite{universal1} with different scales $l/s$, as well as exploring the dynamics in detail.}

The problem we are considering is indeed, the ``cloud-in-cloud" problem. 
The ``cloud-in-cloud" \cite{1986ApJ...304...15B} problem arises when we consider extreme values within specific regions of some field that themselves exhibit high values. 
Specifically, it is about studying extreme events within regions that are already extreme. 
These regions can be thought of as ``clouds" within larger ``clouds". 
Moreover, the details of it may be crucial in determining the nature of the final object that forms. 
This problem is well known, for instance in astrophysics \cite{10.1093/pasj/psaa103}, where it is studied how smaller substructures or ``cloudlets" embedded within larger molecular clouds can successfully collapse and give birth to new stars, despite the external pressures and turbulence exerted by the larger surrounding cloud. A similar consideration applies to halo formation studies \cite{1986ApJ...304...15B}.

%Our aim is 
We aim to consider the ``cloud-in-cloud" problem in the context of PBH formation. 
Having a large peak is already an extraordinary event, and therefore the probability of having two fluctuations overlapped may be even more extraordinary. 
Nevertheless, the relation between the curvature perturbation and the compaction function is non-linear and we find repeated peaks of the compaction function for the typical profile 
with the monochromatic curvature power spectrum. 
Therefore the probability of having such a profile of the compaction function is rather non-trivial. 
In this paper, we only focus on the applicability of existing PBH formation criteria rather than discussing the reality of the overlapping fluctuations.

As already mentioned before, the profile dependence is crucial for determining the threshold and PBH mass, and it is not known how the threshold (defined as the peak value of the critical compaction function) is modified when a secondary peak in the compaction function approaches 
to the first one and/or higher than the first peak. We also want to test wherever the analytical approach of Ref.~\cite{universal1} based on the compaction function shape and its volume average
$\bar{\mathcal{C}}_{c}$ can hold (and in what conditions) in the presence of overlapping two peaks in the compaction function.
% \Blue{compaction functions}. 
In addition, the final PBH mass can be affected by the environment of the fluctuation \cite{Escriva:2021pmf} far away from the central region of the collapse, we would like to test how the PBH mass is affected by the overlapped 
secondary peak in the compaction function
% \Blue{compaction functions} 
in terms of different modulations of the two peaks, and in particular when the second peak can become dominant leading the gravitational collapse. Our work therefore goes further in the direction of studying the profile dependence of the curvature profiles and its effect on the formation of PBHs.

Our paper is organised as follows: In section \ref{sec:basic} we give the basic set-up about how to characterise the initial curvature profiles. In section \ref{sec:threshold_results} we present the numerical results regarding the threshold values for PBH formation. In section \ref{sec:pbh_mass_results} we study in detail the dynamics of PBH formation and the resulting PBH mass for different configurations of the initial conditions. Finally in section \ref{sec:conclusions} we give the conclusions of our work and in the appendixes \ref{sec:details_setup} and \ref{sec:num_convergence}, details of the numerical set-up and some numerical tests to show the convergence of our numerical simulations.

%%%%%%%%%%%%%%%%%%%%%%%%%%%%%%%%%%%%%%%%%%%%%%%%%%%%%%%%%%%%%%%%%%%%%%%%%
\section{Basic set-up}
\label{sec:basic}

We consider the approximation in which the Universe is filled by a perfect fluid with the equation of state $p = w \rho$, with constant $w$ (from now we will consider radiation fluid $w=1/3$), yielding the energy-momentum tensor
\begin{equation}
    T^{\mu \nu} = (p+\rho)u^{\mu}u^{\nu}+p g^{\mu\nu}.
    \label{eq:tensor-energy}
\end{equation}
Here, $p$ is the pressure, $\rho$ is the energy density, $g^{\mu\nu}$ and $u^{\mu}$ are the components of the spacetime metric and of the four-velocity, respectively. Under the assumption of spherical symmetry, the spacetime metric can be written as 
\begin{equation}
\label{eq:2-metricsharp}
    ds^{2}= - A(r,t)^{2} dt^{2}+ B(r,t)^{2} dr^{2}+ R(r,t)^{2} d \Omega^{2},
\end{equation}
with $R(r,t)$ being the  areal radius, $A(r,t)$ the  lapse function, and $d\Omega^{2} \equiv  d\theta^{2} + \sin^{2}(\theta)d\phi^{2}$ the line element of a two-sphere. The definition of the components of the four-velocity $u^{\mu}$ depends on the gauge chosen. In the comoving gauge, we have $u^{t} = 1 / A$ and $u^{i} = 0$ for $i = r$, $\theta$, $\phi$. We use units in which $G_{N} = c = 1$ (geometrised units).

Under this approach, we solve numerically Misner-Sharp equations following \cite{escriva_solo,Escriva:2020tak} (see the references for details and the appendix \ref{sec:details_setup}). The initial conditions are implemented following the gradient expansion approach \cite{1996CQGra..13..705N,1996PThPh..95..295T} when the cosmological fluctuations lie at much larger scales than the cosmological horizon. Then, the spacetime metric at super-horizon scales is defined as\footnote{Notice that we can also define the curvature fluctuations in terms of $\zeta$, the relation between $\zeta$ and $K$ is given by a change of coordinates \cite{refrencia-extra-jaume}.}
\begin{equation}
\label{eq:2-FLRWmetric5}
    d s^{2}= -d t^{2}+a^{2}( t )\mspace{-1.5mu}\left[\frac{d r^{2}}{1 - K( r )r^{2}}+r^{2}d\Omega^{2}\right].
\end{equation}
The initial conditions to solve numerically Misner-Sharp equations are given following \cite{Polnarev:2006aa,2012JCAP...09..027P,refrencia-extra-jaume} at the first order in gradient expansion, considering that the curvature fluctuations are initially frozen at super-horizon scales (see the appendix \ref{sec:details_setup} for details). 
%For instance, the density contrast of the fluctuation at super-horizon scales is given by,
%\begin{equation}
    %\frac{\delta \rho}{\rho_b} = \frac{2}{3}\left( \frac{1}{a H} %\right)^{2}\left[ K(r)+\frac{r}{3}K'(r) \right]
%\end{equation}
A well known useful strength estimator to characterise the formation of a PBH is the compaction function $\mathcal{C}$, introduced in \cite{Shibata:1999zs}. It is defined as\footnote{In other references is commonly defined Eq.\eqref{eq:compact_function} without the factor $2$.} twice the mass excess inside a given areal radius $R$ in the comoving gauge (see \cite{Harada:2023ffo} for its gauge dependence),
\begin{equation}
    \mathcal{C}\coloneqq 2\frac{M_{\rm MS}-M_{b}}{R},
    \label{eq:compact_function}
\end{equation}
where $M_\MS \equiv 4\pi\int_0^R\rho\tilde{R}^2\dd{\tilde{R}}$ is the Misner--Sharp mass (which takes into account the kinetic and potential energy) and $M_b$ is the mass expected in the FLRW background, defined as $M_b=4 \pi \rho_{b}R^{3}/3$. $\rho$ is the energy density of the 
%full 
fluid, while $\rho_b$ is that of the FLRW background, which evolves as $\rho_b = \rho_{b,0}(t/t_{0})^{-2}$. 
As already noticed from time ago by Shibata $\&$ Sasaki \cite{Shibata:1999zs} and explored further in \cite{refrencia-extra-jaume,Musco:2018rwt,universal1,Escriva:2020tak,2021PhRvD.103f3538M,Harada:2023ffo}, the compaction function's peak is a useful estimator to characterise fluctuations that will lead to black hole formation. The criteria we follow is that if the initial compaction function contains a peak $\mathcal{C}(r_m)$ (being $r_m$ the location of that peak) bigger than a given threshold, then the gravity pull will be stronger than the pressure gradients, and lead to collapsing and forming a black hole. On the opposite side, the fluctuations will be dispersed on the FLRW background.
%where the compaction function measure the mass excess , given by\footnote{In other references is commonly defined Eq.\eqref{eq:compact_function} without the factor $2$.}
Hereafter the critical values and profiles of the compaction function will be indicated by the subscript $c$, so that 
the compaction function with the amplitude of the threshold value is expressed as $\mathcal C_c(r)$. 

Therefore, we will characterise the amplitude of the fluctuations as the peak value of the compaction function. Specifically, at super-horizon scales, the compaction function is a time-independent quantity, whose expression in the comoving gauge was found in \cite{refrencia-extra-jaume} (see their Eqs.~(6.34) and (6.35)):
\begin{equation}
    \mathcal{C} = \frac{2}{3} K(r) r^{2},
    \label{eq:compact_K}
\end{equation}
where the factor $2/3$ comes from the fact that $w=1/3$. 

It was found in \cite{universal1} that the averaged critical compaction function defined as 
\begin{equation}
\label{3_Cbar}
\bar{\mathcal{C}}_{c}\equiv
 \frac{3}{ r_{m}^3 }\int_{0}^{ r_{m}} \mathcal{C}_{c}(\tilde{r}) \tilde{r}^2 d\tilde{r}\ ,
\end{equation}
is approximately equal to $\bar{\mathcal{C}_{c}}=2/5$ in radiation dominated-Universe (for a generalisation with other equations of state see \cite{Escriva:2020tak}) and mostly universal, independent of the curvature profile considered within a deviation $\mathcal{O}(2\%)$. 

Another way to characterise the critical compaction function is to use the $p$ parameter (introduced 
%for the first time in the literature 
in \cite{universal1}\footnote{Notice that 
%there is defined as 
the corresponding parameters are defined with "$q$" letter for isolated peaks. }
) 
%associated with each peak of $\mathcal{C}_{j}$ can be 
computed as:
\begin{equation}
    %p_{j} = -\frac{\mathcal{C}''(r_{m,j}) r^2_{m,j}}{4 \mathcal{C}(r_{m,j})}.
    p = -\frac{\mathcal{C}''(r_m) r^2_{m}}{4 \mathcal{C}(r_{m})}.
\end{equation}
%Different $p$-values allow a modulation on the shape around the compaction function's peak: large 
Larger values of $p$ %$p$ 
%values 
correspond to 
%a 
sharp profiles around the compaction function peak, 
%which 
and give the larger thresholds (being $2/3$ the maximum). 
On the other hand, small $p$ corresponds to a broad shape of $\mathcal{C}$, which gives the smaller threshold (being $0.4$ the minimum). 
As found in \cite{universal1} different profiles/initial conditions characterised by the same $p$ lead to the same threshold value up to a deviation of $\mathcal{O}(2\%)$, which means that the threshold of the gravitational collapse mainly depends on the shape around the compaction function peaks. 

Let us introduce the following specific profile of the compaction function $\mathcal C_0$ characterised by the parameter $p_0$ and $r_0$:
% where $\mathcal{C}_{j}$ is equal to:
\begin{equation}
\mathcal{C}_{0}(r, \delta,p_0,r_{0}) = \delta \left(\frac{r}{r_{0}}\right)^{2} e^{\frac{1}{p_0}\left[ 1-\left( \frac{r}{r_0} \right)^{2 p_0}  \right]} ,
\label{eq:basis_C_exp}
\end{equation}
where the parameters $p_0$ and $r_0$ are equal to $p$ and $r_m$ in this profile, respectively.%
%which are exponential compaction function profiles parameterised by the parameter $p_0$
\footnote{In \cite{Escriva:2020tak} it has been considered polynomial profiles instead of exponential. The advantage is that polynomial profiles fulfill regularity conditions at the centre $r=0$ for any $p$, which is not the case for exponential where regularity conditions are violated for $p<0.5$ and turns this kind of profile (and therefore the initial condition) to be physically inconsistent. Nevertheless, we don't have this issue in this work since we consider $p>0.5$ always, and exponential profiles decay faster than polynomials, allowing a better modulation of overlapped compaction functions}. 
%So basically $K_{\rm total} = K_{1}+K_{2}$ with $K_{j}=3 \mathcal{C}_{j}/(2 r^2)$. 
The use of this profile and the value $\bar{\mathcal{C}_{c}}=2/5$
allow us to build an analytical formula given by:
\begin{equation}
\delta_{c}(p) = \frac{4}{15} e^{-\frac{1}{p}}\frac{p^{1-\frac{5}{2p}}}{\Gamma\left(\frac{5}{2p}\right)-\Gamma\left(\frac{5}{2 p},\frac{1}{p}\right)}.
\label{eq:threshold_analit}
\end{equation}
% using the fact that $\bar{\mathcal{C}_{c}}=2/5$. 
Although this formula is derived through the specific profile of the compaction function, 
since the same $p$ leads to the approximately same threshold value independent of the specific profile, 
this formula is expected to apply to any profile of the inhomogeneity.

Let's now consider the situation where we have two overlapped peaks in the compaction function
% compaction functions. 
We introduce fluctuations that are overlapped to another one on much larger scales. We therefore define two fluctuations $K_1$ and $K_2$. 
The ``total" compaction function $\mathcal{C}$ is defined as,
\begin{equation}
\mathcal{C}(r) = \mathcal{C}_{1}(r, \delta_{1},p_{1},r_{1})+ \mathcal{C}_{2}(r,\delta_{2},p_{2},r_{2}), 
\label{eq:4_dos_torres}
\end{equation}
where $\mathcal C_i(r, \delta_{i},p_{i},r_{i})=\mathcal C_0(r, \delta_{i},p_{i},r_{i})$ for $i=1$ and $2$. 

\begin{figure}[h]
\centering
\includegraphics[width=2.5 in]{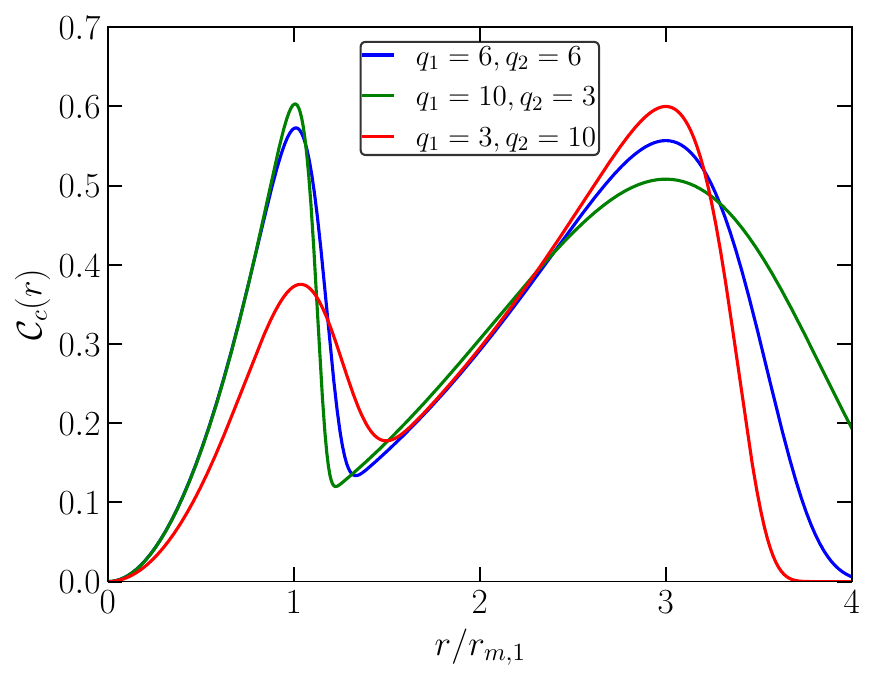}
\includegraphics[width=2.5 in]{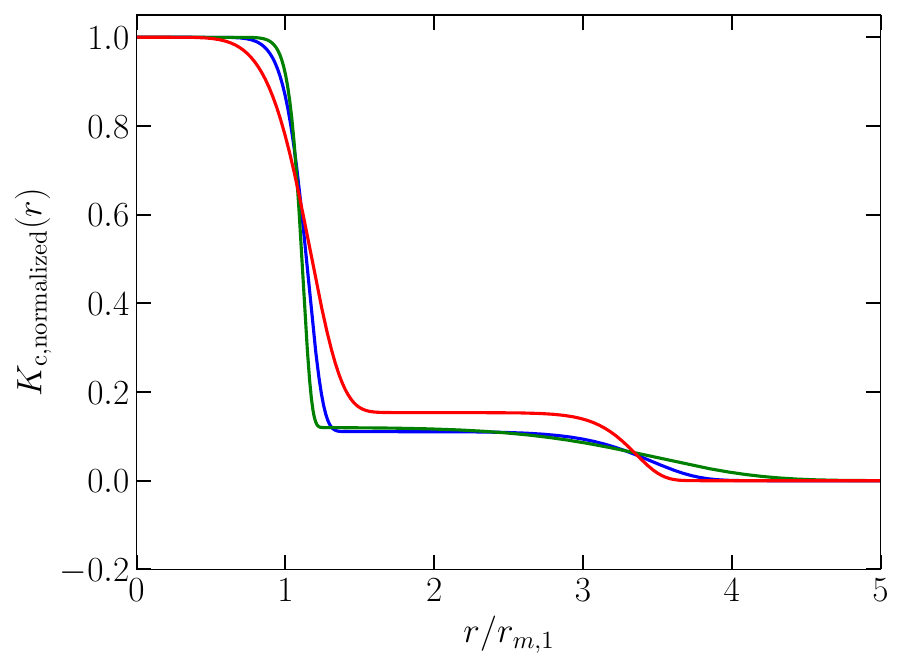}
\includegraphics[width=2.5 in]{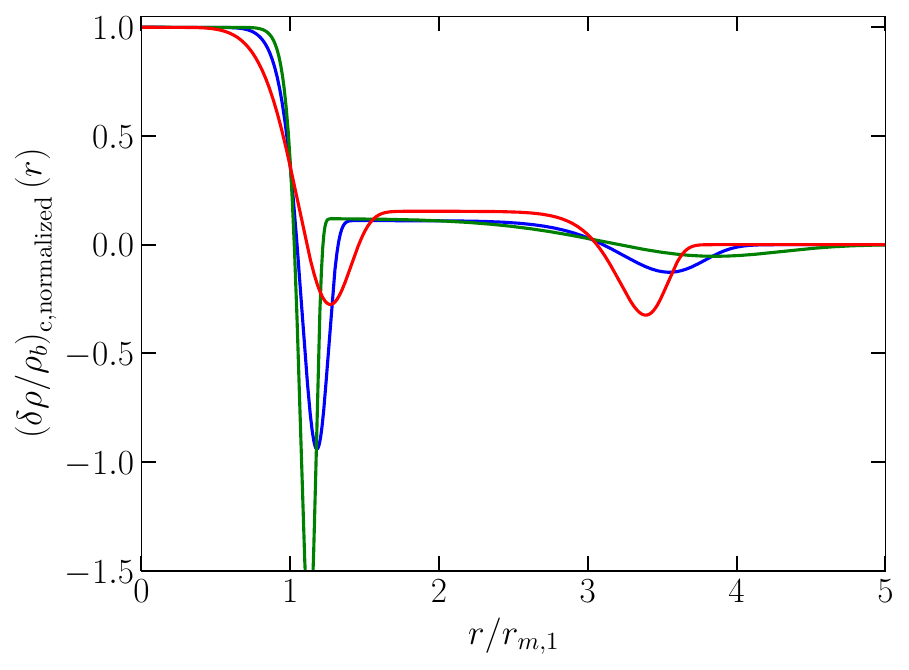}
\caption{Examples of critical compaction function (top-left), critical curvature (top-right) and critical density contrast (bottom) for three different configurations of $q_{j}$ with $\beta=3$. The values for the density contrast and the curvature have been normalised to the peak value ($r=0$).}
\label{fig:C_two_peaks11}
\end{figure}

We define $r_{m,j}$ as the location of the peaks ($j=1,2$) of the total compaction $\mathcal{C}(r)$ and $\beta \equiv r_{m,2}/r_{m,1}$ its ratio. Notice that  $r_{m,j} \approx r_{j}$, but not equal since there will be an overlap between 
$\mathcal{C}_{1}$ and $\mathcal{C}_{2}$. 
% the two \Blue{compaction functions $\mathcal{C}_{j}$}. 
Indeed, if we only consider one single curvature fluctuation given by $K=K_{j}:=3 \mathcal{C}_{j}/(2 r^2)$ then $r_{m,j} = r_{j}$. 
Eqs.\eqref{eq:threshold_analit} is then generalised by using the effective $q_j$ that takes into account the $r_{m,j}$ and the total shape of the compaction function, 
\begin{equation}
q_{j} = -\frac{\mathcal{C}''(r_{m,j}) r^2_{m,j}}{4 \mathcal{C}(r_{m,j})}.
\end{equation}
Then the amplitude of each peak will be given by $\mathcal{C}(r_{m,1})$ and $\mathcal{C}(r_{m,2})$. 
Some examples of the profiles considered for different parameters can be found in Fig.~\ref{fig:C_two_peaks11}.

%%%%%%%%%%%%%%%%%
%%%%%%%%%%%%%%%%%%%
%%%%%%%%%%%%%%%%%%%

\begin{figure}[h]
\centering
\includegraphics[width=3.5 in]{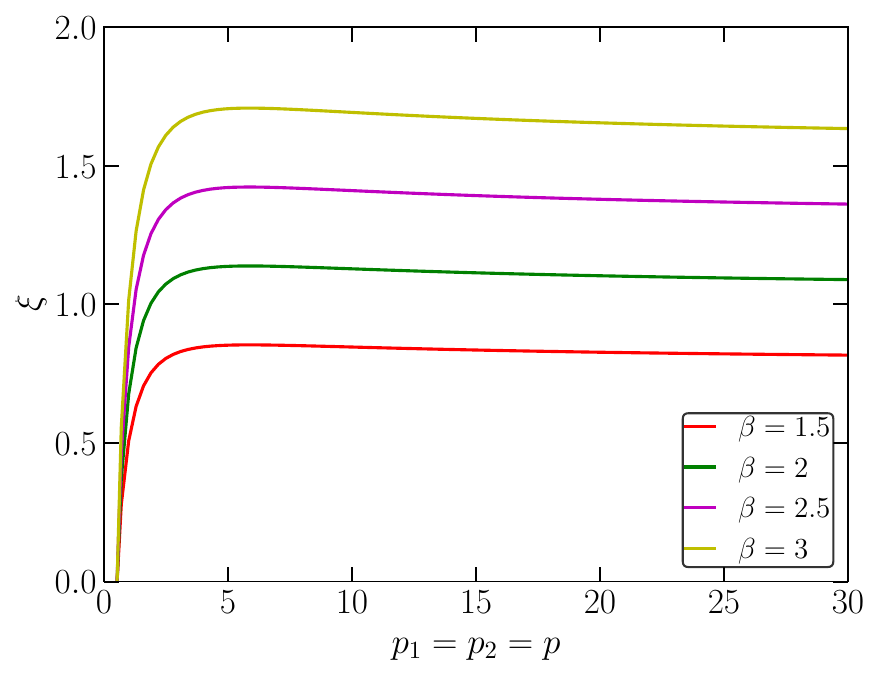}
\caption{Values of $\xi$ from Eq.\eqref{eq:xi_eq} for the different configurations $p_{1}=p_{2}=p$.}
\label{fig:sigmas_cosas}
\end{figure}

To quantify the degree of overlapping between  
$\mathcal{C}_{1}$ and $\mathcal{C}_{2}$, we define the following quantity 
\begin{equation}
    \xi = \frac{r_{2}-\sigma_{2}}{r_{1}+\sigma_{1}},
    \label{eq:xi_eq}
\end{equation}
where we define $\sigma_{j}$ associated to 
% each \Blue{compaction function $\mathcal{C}_{j}$} 
$\mathcal{C}_{j}$ as:
\begin{equation}
    \sigma_j = \frac{\sqrt{\int_{0}^{\infty} \mathcal{C}_{j}(r)(r_{j}-r)^2 dr}}{\sqrt{\int_{0}^{\infty} \mathcal{C}_{j}(r)dr}}.
\end{equation}
The integrals can be done analytically introducing Eq.\eqref{eq:basis_C_exp}, and we obtain:
\begin{equation}
\sigma^2_{j}= \frac{3}{2}r^{2}_{j}\frac{\Gamma(3/2p_j)-2p^{1/2 p_j}_j\Gamma(2/p_j)+p^{1/p_j}_j\Gamma(5/2 p_j)}{p_{j} \Gamma(1+3/2p_j)}.
\end{equation}
In Fig.~\ref{fig:sigmas_cosas} we show the variation of the parameter $\xi$ for different configurations of $p_1=p_2=p$ and $\beta$. When $\xi>1$ the two peaks in the compaction function associated with
% compaction functions 
$\mathcal{C}_1$, $\mathcal{C}_2$ %becomes 
are decoupled, in the opposite case when $\xi \rightarrow 0$ are completely coupled. $\xi$ saturates for sufficiently large values of $p_{j}$ and $\beta$. Notice that this is consistent with having completely overlapped profiles when $p_{j} \rightarrow 0$ since the compaction function becomes constant and homogeneous. It already indicates to us that for $\beta \approx 3$ with $p_{j}> 3 $ the initial configuration of overlapped 
peaks in the compaction function
% \Blue{compaction functions} 
can be approximately considered as isolated. 
    That is, %Notice that 
    for sufficiently small $q_{j}$ and/or $\beta$, the 
    % \Blue{compaction functions} 
    peaks in the compaction function
    are completely overlapped and 
    can not be differentiated (only a single peak). 
    %For as}.     
Since we are interested in the distinct 
peaks in the compaction function
% overlapping \Blue{compaction functions} 
of two different scales, we restrict ourselves to 
the situation in which two scales are isolated with the parameters $q_i\geq3$ and $\beta\geq2$ hereafter.

We may define now the time $t_{H,j}$ that the length-scale $r_{m,j}$ takes to reenter the cosmological horizon given by
\begin{equation}
\label{eq:time_horizon_crosing}
t_{H,j} = t_0 \left(a_0 \frac{r_{m,j}}{R_{H}(t_0)}\right)^{2},
\end{equation}
where $a_0$, and $t_0$ are gauge quantities that we set to one. 
Moreover, we define the mass of the cosmological horizon $M_{H}=1/(2 H)$, which at the time $t_{H,j}$ is given by $M_{H}(t_{H,j})=t_{H,j}$ and $R_H(t_0)=1/H_0$.

\section{Threshold values}
\label{sec:threshold_results}
To explore the effect of the overlapped peaks in the compaction function 
% \Blue{compaction functions} 
on the necessary initial condition for PBH formation, we run a numerical bisection over different trial values $\delta_{1}$, $\delta_{2}$ fixing $q_{1}$, $q_{2}$ and the ratio between the length-scales $\beta \equiv r_{m,2}/r_{m,1}$ to find the critical value of the two peaks, i.e, $\mathcal{C}_{c}(r_{m,1})$ and $\mathcal{C}_{c}(r_{m,2})$ \footnote{We make the numerical bisection to obtain the threshold values with a resolution of $\mathcal{O}(0.1\%)$ in percentage, which means an absolute resolution of $\mathcal{O}( 10^{-4})$.}. Notice that due to the overlapping, 
% between the \Blue{compaction functions}, 
we have to tune the parameter $p_i$ so that a fixed value of $q_{j}$ will be realised for each numerical iteration.

\begin{figure}[h]
\centering
\includegraphics[width=3. in]{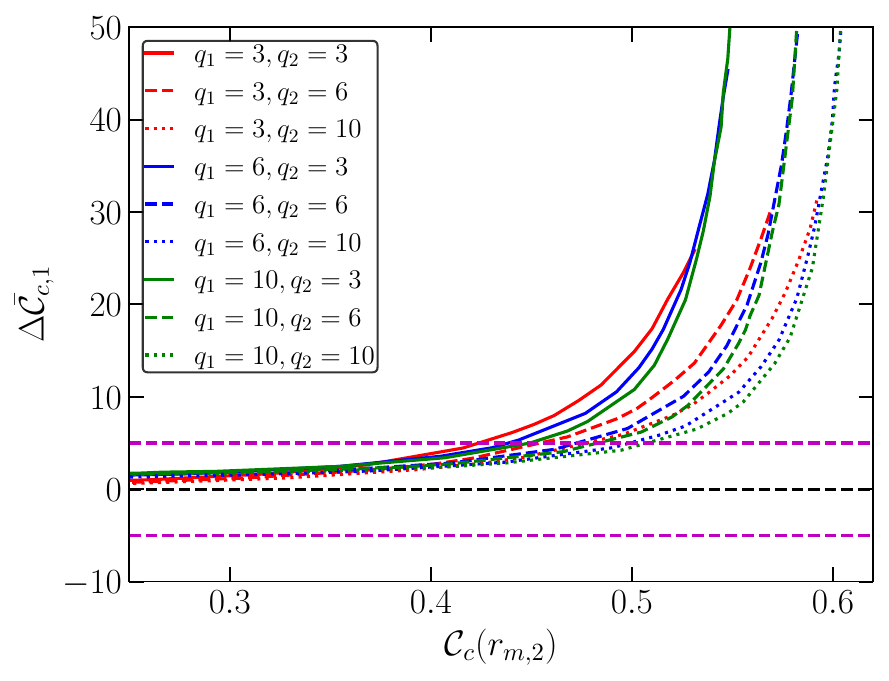}
\includegraphics[width=3. in]{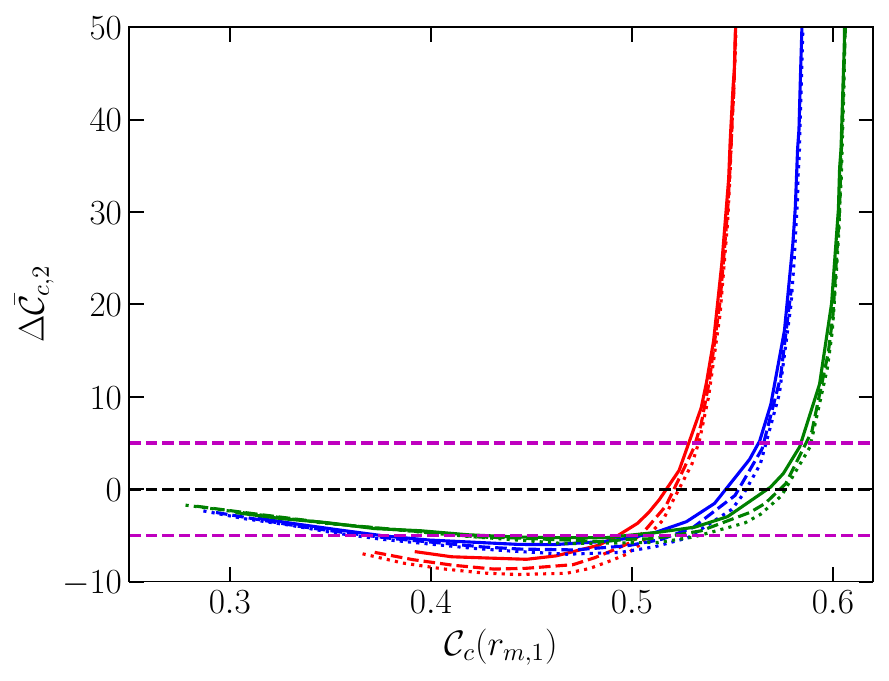}
\caption{Relative deviation of the averaged critical compaction function Eq.\eqref{eq:percentatge_averaged}, computed for the first peak (left-panel), and up to the second peak (right-panel). The dashed magenta lines corresponds to a deviation of $\pm 5\%$.}
\label{fig:averaged_2}
\end{figure}

With such values, we can compare the numerical thresholds from the simulations with the case of having an isolated compaction function with $p_{j}=q_{j}$ and $\mathcal{C}_{\rm c,j}(r_j) \equiv \delta_{c,j}$. This corresponds to the peak value for the critical compaction function when we only consider a single curvature fluctuation $K_{j}$ (isolated fluctuation) and is computed following Eq.\eqref{eq:threshold_analit} with the effective $q_{j}$ associated to $\mathcal{C}_{c}(r_{m,j})$. We make this comparison by computing the following relative deviation,
%without considering the overlap between the fluctuations. We define this deviation in percentage as,
%
\begin{equation}
\label{eq:percentatge}
\Delta \mathcal{C}_{c,j} (\%) = 10^{2} \cdot \frac{\mathcal{C}_{c}(r_{m,j})-\delta_{c,j}}{\mathcal{C}_{c}(r_{m,j})}.
\end{equation}
We have also computed the averaged critical compaction function $\bar{\mathcal{C}}_{c,j}$ integrated up to the first ($j=1$) and second peak ($j=2$), which is shown in Fig.~\ref{fig:averaged_2}. To compare with the value found in the case of having isolated fluctuations $\bar{\mathcal{C}_c}=2/5$ we define
\begin{equation}
\label{eq:percentatge_averaged}
\Delta \bar{\mathcal{C}}_{c,j}(\%)  = 10^{2} \cdot \frac{ \bar{\mathcal{C}}_{c,j}-2/5}{2/5}.
\end{equation}

First, let us check the relative deviation of the averaged compaction function $\bar{\mathcal C}$. When the gravitational collapse is lead by the first peak of the compaction function (having $\mathcal{C}_{c}(r_{m,2})$ well under threshold), the value of $\bar{\mathcal C}$ up to the first peak is well approximated by the value $2/5$. 
When the second peak is near the threshold value, 
 $\bar{\mathcal{C}}_{c}$ becomes much larger than the value $2/5$. 
% This starts to differ when the second peak is near the threshold value, which makes the gravitational collapse 
%to be 
% overlapped between the two \Blue{compaction functions} being 
%the averaged 
% $\bar{\mathcal{C}}_{c}$ larger than expected. 
The same behaviour is shown in the right panel of Fig.~\ref{fig:averaged_2} but now considering the second peak. 
Although the differences between the profiles considered are not substantial, 
%Still, 
we can identify in the right panel that the deviation $\Delta \bar{\mathcal{C}}_{c,2}$ becomes larger when $q_{1} =3$ in comparison with other cases of $q_{1}$, since the compaction function profile is broader.
We also note that the right and left panels in Fig.~\ref{fig:averaged_2} are qualitatively different. 
More specifically, we find $\Delta \bar{\mathcal{C}}_{c,1}>0$ and $\Delta \bar{\mathcal{C}}_{c,2}<0$ in those relevant regions.  
Moreover, in the right panel, we can find a qualitative difference in the behaviour of $\Delta \bar{\mathcal{C}}_{c,2}$ depending on the parameter $q_1$.
These facts indicate that the averaged critical compaction function $\bar{\mathcal C}_c$ could become sensitive to the structure inside the radius $r_m$ for some specific curvature profiles, in this case due to the effect of the first peak in the gravitational collapse. Although the values of $\Delta \bar{\mathcal{C}}_{c,i}$ are not significantly large in our settings, 
this sensitivity may spoil the criterion with $\bar{\mathcal C}$ as was already observed in \cite{2022JCAP...05..012E}.

%%%%%%%%%%%%%%%%%% 

\begin{figure}[h]
\centering
\includegraphics[width=3. in]{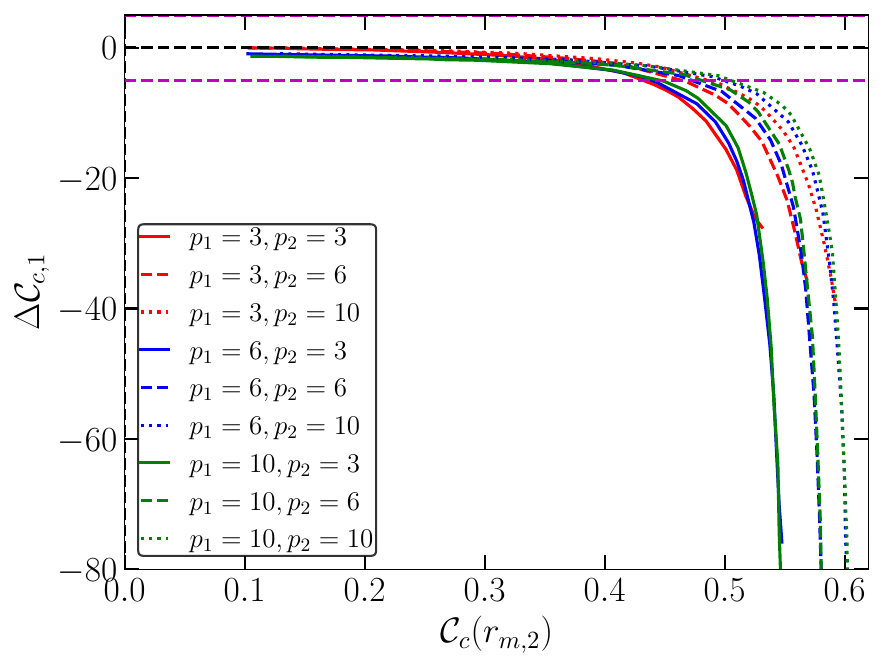}
\includegraphics[width=3. in]{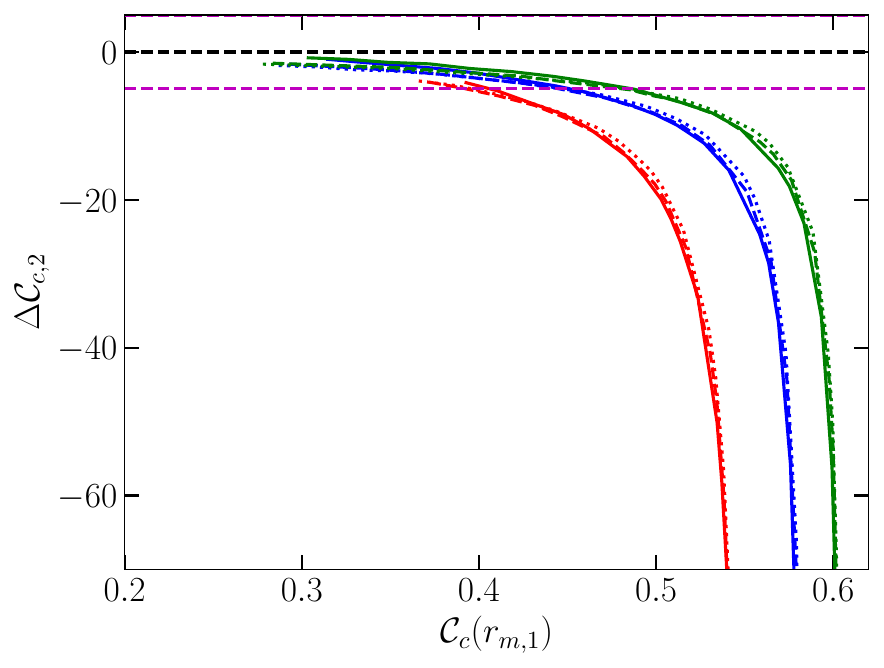}
\includegraphics[width=3. in]{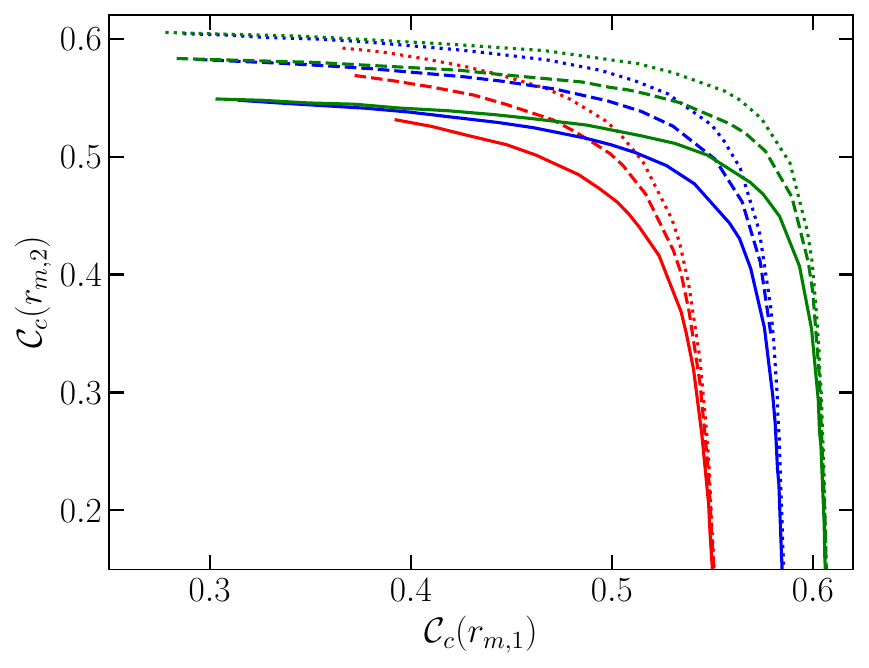}
\includegraphics[width=3. in]{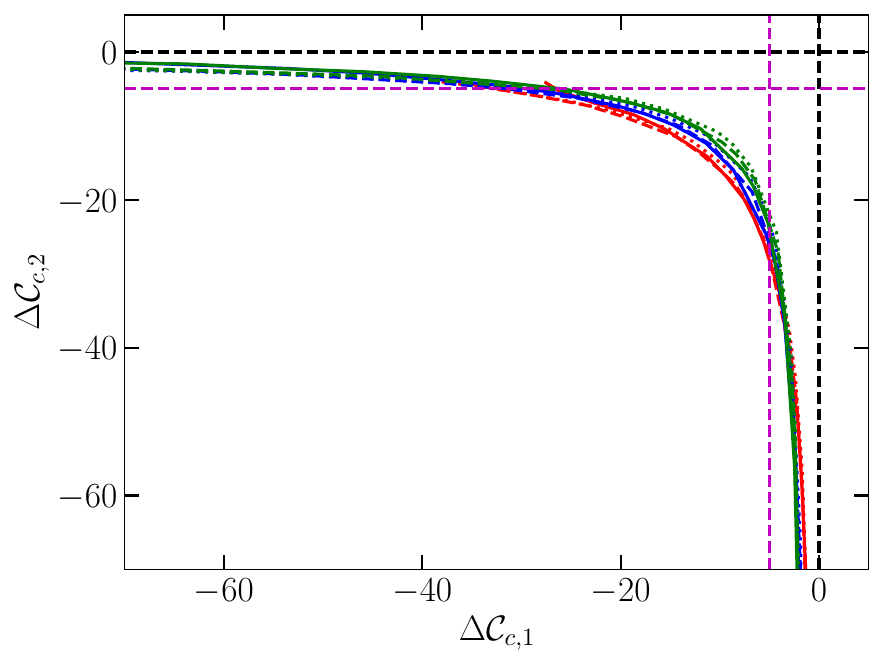}
\caption{Top panels: Eq.\eqref{eq:percentatge} computed for the first peak and compared with the second peak (left panel) and for the vice-versa (right panel). Bottom-left panel: Phase diagram of the two thresholds ($\mathcal{C}_c(r_{m,1})$ , $\mathcal{C}_c(r_{m,2})$). Bottom-right panel: Relative deviation of Eq.\eqref{eq:percentatge} compared between both peaks.}
\label{fig:resultados_beta2}
\end{figure}

Next, let us check the value of $\Delta \mathcal{C}_{c,j}$. 
% Instead, i
In Fig.~\ref{fig:resultados_beta2},  we show the numerical results for the thresholds values $\mathcal{C}_{c}(r_{m,j})$ and 
its relative deviation following Eq.\eqref{eq:percentatge}, choosing different configurations of $q_j$ fixing the ratio $\beta=2$.
In the top panel, the thresholds $\mathcal{C}_{c}(r_{m,j})$ are shown compared with $\Delta \mathcal{C}_{c,j}$. 
When $\Delta \mathcal{C}_{c,1} \lesssim -50$ the first peak is well under the threshold $\delta_{c,1}$ and therefore, the critical threshold value of the secondary peak is mainly given by $\mathcal{C}_{c}(r_{m,2}) \rightarrow \delta_{c,2}$. In this case, the second peak will lead the gravitational collapse. The analogous behaviour is found when $\Delta \mathcal{C}_{c,2} \lesssim -50$, the second peak is well under the threshold $\delta_{c,2}$ and therefore, the critical value of the first peak is given by $\mathcal{C}_{c}(r_{m,1}) \rightarrow \delta_{c,1}$, which will lead the gravitational collapse.

Possibly, the most interesting case is when the two peaks are under the thresholds $\delta_{c,j}$, in other words, $\mathcal{C}_{c}(r_{m,1})<\delta_{c,1}$ and $\mathcal{C}_{c}(r_{m,2})<\delta_{c,2}$. In this case, the overlapped 
peaks in the compaction function
% \Blue{compaction function} 
will contribute to the gravitational collapse of the shorter $\mathcal{C}_{1}$, lowering the threshold values by a few percentages compared with the ideal case of isolated 
peaks
% \Blue{compaction functions} 
(see bottom-left panel of Fig.~\ref{fig:resultados_beta2}). Notice that it depends on the profile considered. For large $q_{j}$, the shape around the compaction function peak will be sharper with a small mass excess, and therefore, the overlapping between the two peaks will be smaller (compare the green lines with the red ones). 
%the deviation between the numerical results and the analytical formula $\delta_{c,j}$ of Eq.\eqref{eq:threshold_analit} lies within a $5\%$ (magenta dashed lines). 
Finally, the phase diagram of $\mathcal{C}_{c}(r_{m,1})$ and $\mathcal{C}_{c}(r_{m,2})$ is shown in the bottom-left panel. We find $\mathcal{C}_{c}(r_{m,2}) \ll \delta_{c,2} \Rightarrow \mathcal{C}_{c}(r_{m,1}) \rightarrow \delta_{c,1}$ and $\mathcal{C}_{c}(r_{m,1}) \ll \delta_{c,1} \Rightarrow \mathcal{C}_{c}(r_{m,2}) \rightarrow \delta_{c,2}$.

%Notice that the approach of using the analytical formula $\delta_{c}(q)$ seems to be more robust in comparison with the averaged $\mathcal{C}_c$ when comparing the numerical results with the analytical approaches for the particular cases tested. More specifically, we do not find any qualitative difference between the behaviours of $\Delta \mathcal{C}_{c,1}$ and $\Delta \mathcal{C}_{c,2}$ unlike those in the case of the averaged compaction function, where the profile dependence seems more important, as can be noticed by comparing the Fig.4 (right panel) with Fig.5 (top-right panel) for the cases $q_1=3$ (red lines).

Notice that, for the particular cases tested, the approach of using the analytical formula $\delta_{c}(q)$ seems to be more robust in comparison with the averaged $\mathcal{C}_c$ when comparing the numerical results with the analytical approaches. More specifically, we do not find any qualitative difference between the behaviours of $\Delta \mathcal{C}_{c,1}$ and $\Delta \mathcal{C}_{c,2}$ unlike those in the case of the averaged compaction function, where the profile dependence seems more important, as can be noticed by comparing the Fig.~\ref{fig:averaged_2} (right panel) with Fig.~\ref{fig:resultados_beta2} (top-right panel) for the cases $q_1=3$ (red lines).

\begin{figure}[h]
\centering
\includegraphics[width=3. in]{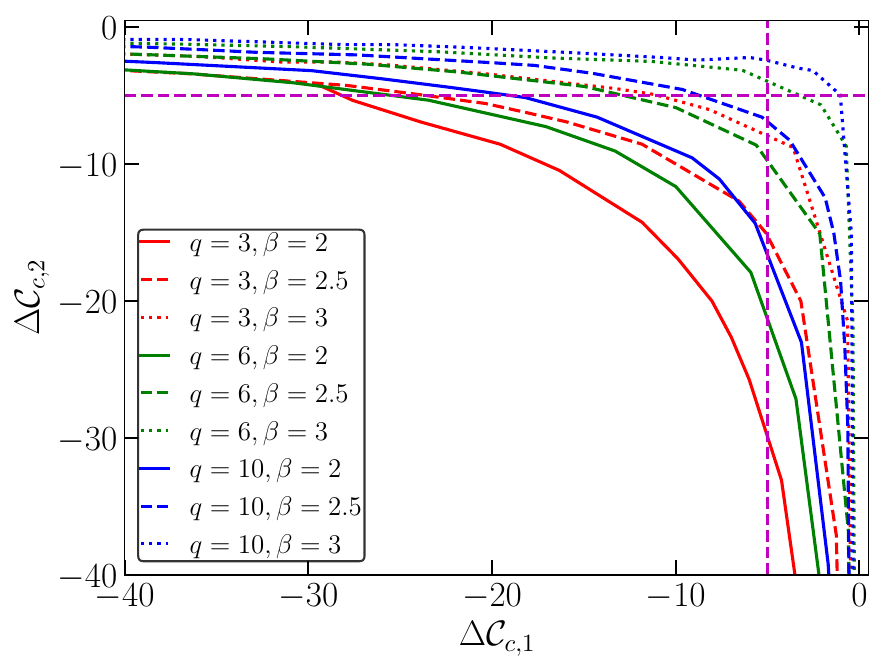}
\includegraphics[width=3. in]{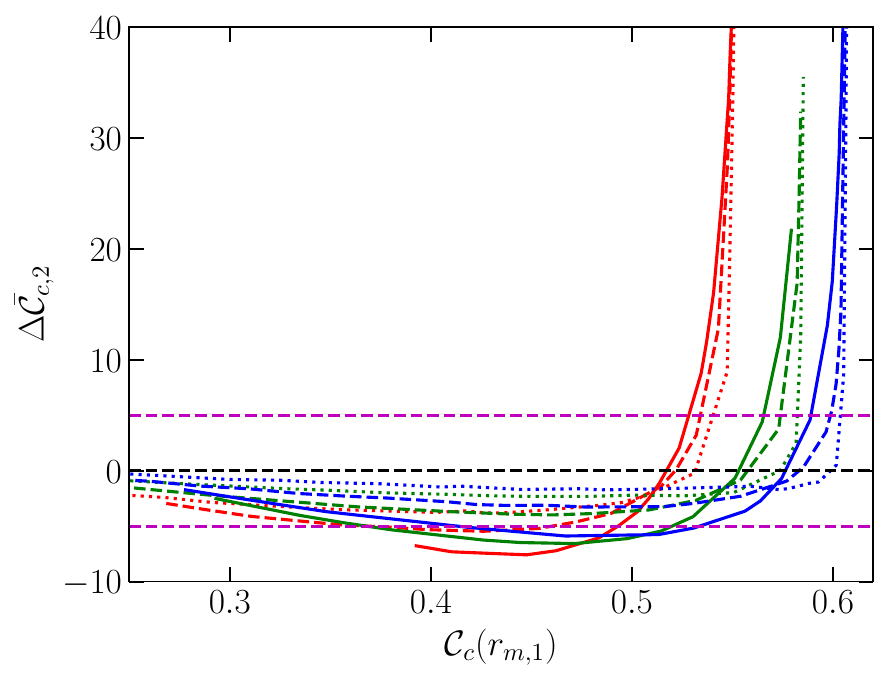}
\includegraphics[width=3. in]{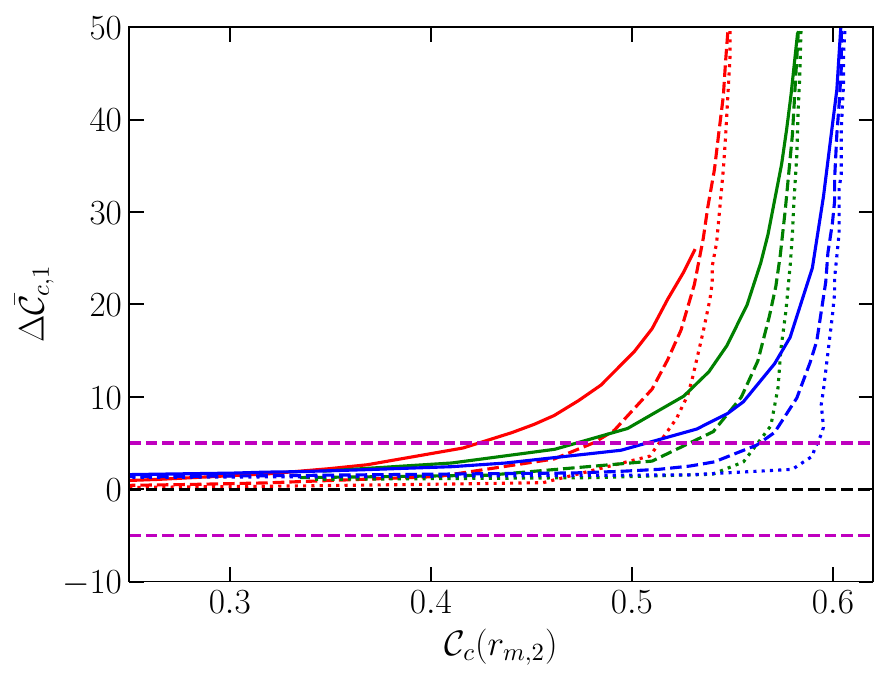}
\includegraphics[width=3. in]{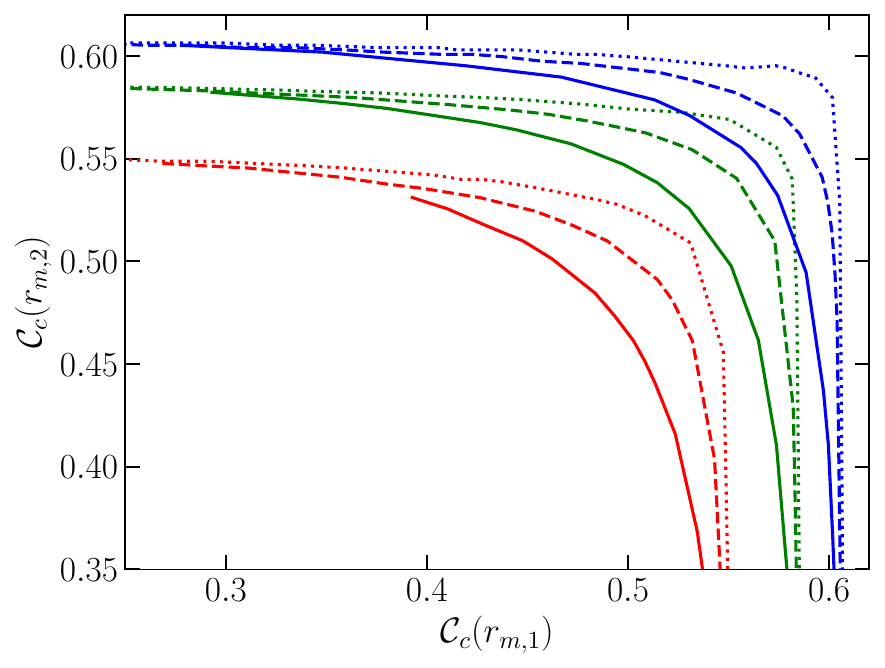}
\caption{Top-left panel: Relative deviation of Eq.\eqref{eq:percentatge} compared between both peaks. Top-right panel: Relative deviation of the averaged critical compaction function Eq.\eqref{eq:percentatge_averaged} for the second peak. Bottom-left panel: the same as in the previous case but for the first peak. Bottom-right panel: Threshold values of the two peaks.}
\label{fig:dependence_beta}
\end{figure}

% On the other hand, 
In Fig.~\ref{fig:dependence_beta}, we show the dependence of the previous quantities in terms of the distance $\beta$. For simplicity, we now consider the cases where $q_{1}=q_{2}$. 
% Both \Blue{compaction functions} 
Two peaks in the compaction function
become more decoupled when increasing $\beta$, making the gravitational collapse dominated by one of the two peaks. This makes the relative deviations smaller when increasing $\beta$, in comparison with Figs.~\ref{fig:averaged_2} and \ref{fig:resultados_beta2}. 

\begin{figure}[h]
\centering
\includegraphics[width=3. in]{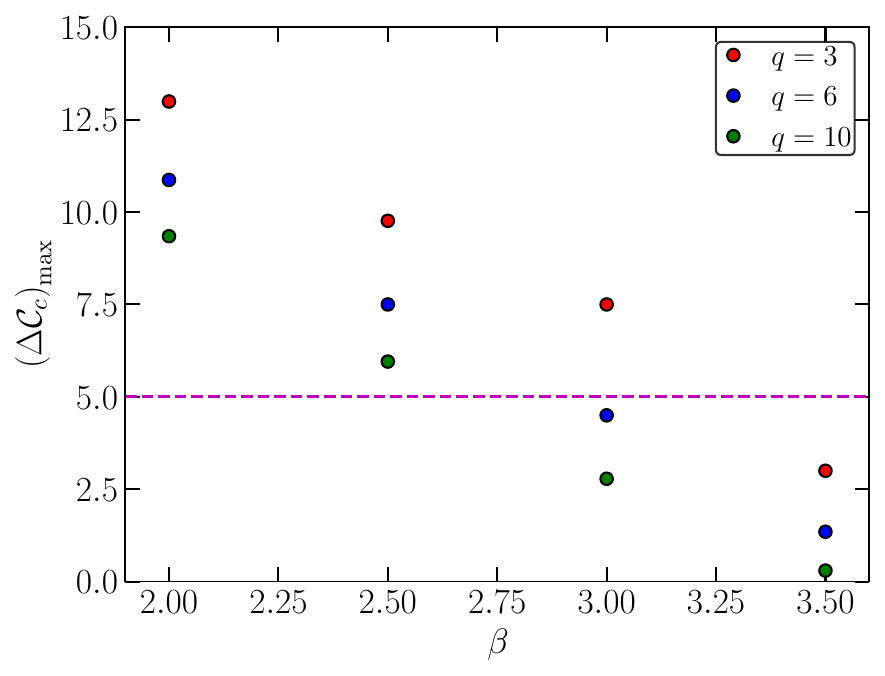}
\caption{Maximum relative deviation Eq.\eqref{eq:percentatge} in terms of $\beta$ for different profiles.}
\label{fig:dependence_beta2}
\end{figure}

To quantify the extent of the effect of the overlap between the two 
peaks in the compaction function, 
% \Blue{compaction functions}, 
we 
use the maximum distance between the origin and the curves of $\Delta \mathcal{C}_{c,j}$ (see the top-left panel of Fig.~\ref{fig:dependence_beta}). 
Then, as shown in Fig.~\ref{fig:dependence_beta2}, the maximum deviation 
is around $10\%$ at worst in our settings. 
The deviation almost linearly depends on the parameter $\beta$ and 
reduces to smaller values than $5\%$ 
for $\beta\gtrsim3$ depending on the parameter $q$. 
% when the two peaks are sufficiently decoupled. This behaviour depends on the profile considered. 
As expected, when the 
two scales of peaks
% \Blue{compaction functions} 
are sufficiently separated, 
the threshold is well approximated by the analytic formula for each isolated peak. Our results show that even in the presence of secondary peaks in the compaction function as a function of a radial coordinate, if those peaks are sufficiently separated and decoupled, the threshold of the gravitational collapse basically depends on the shape around the critical compaction function as found in \cite{universal1}, irrespectively of the new scale introduced.

\section{Dynamical evolution and PBH mass}
\label{sec:pbh_mass_results}

\subsection{Dynamics of fluctuations}
\begin{figure}[h]
\centering
\includegraphics[width=2.5 in]{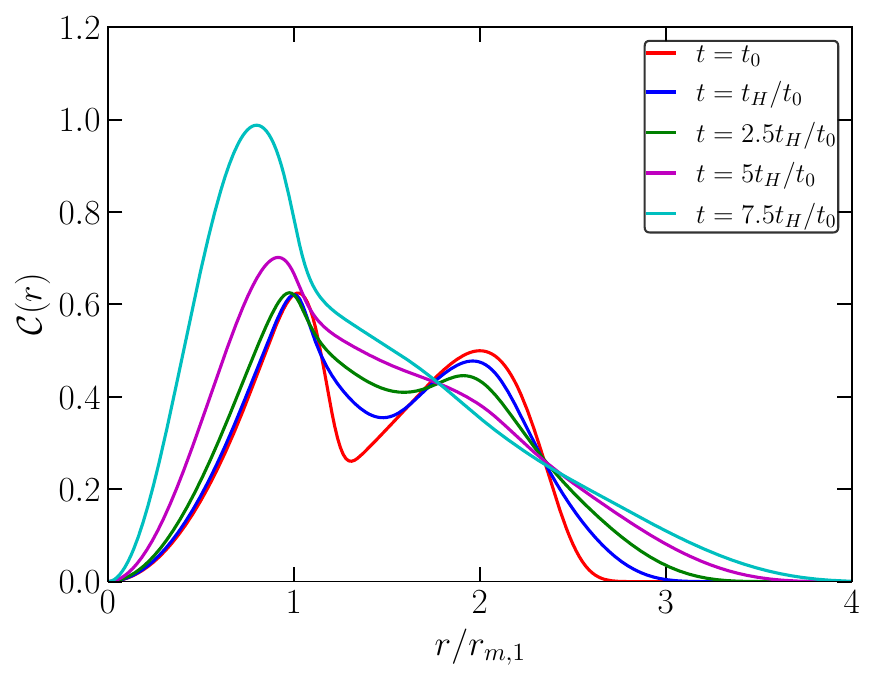}
\includegraphics[width=2.5 in]{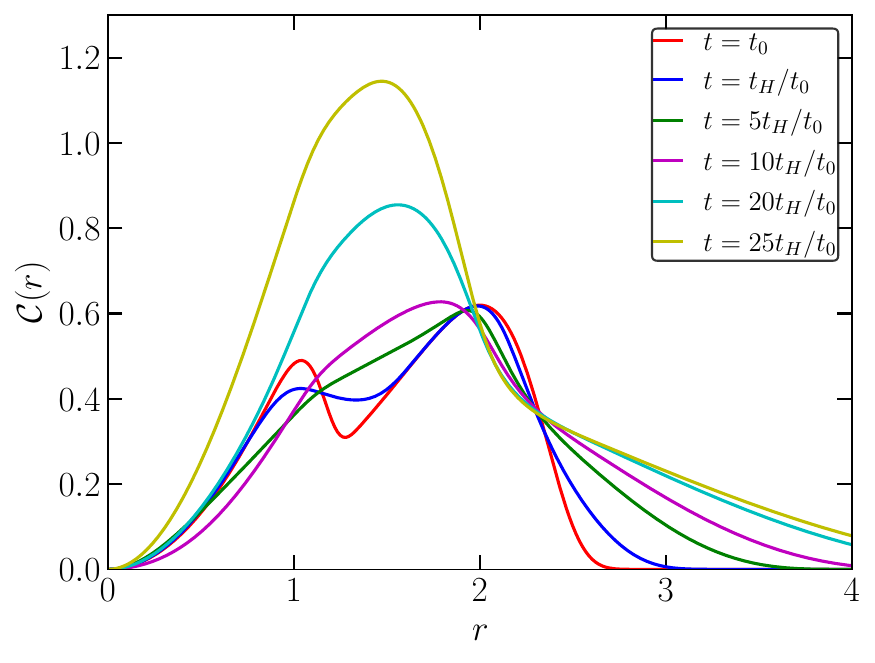}
\includegraphics[width=2.5 in]{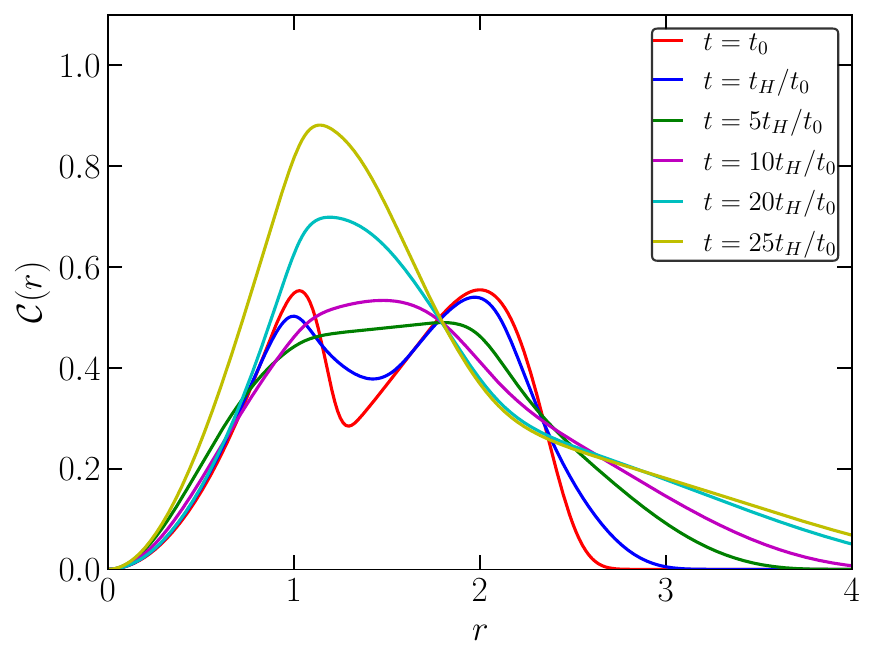}
\caption{Snapshots of the evolution of the compaction function. Top-left panel: Case with $\delta_1 = 0.475$ and $\delta_{2}= 0.5$, yielding to $\mathcal{C}(r_{m,1})= 0.626$, $\mathcal{C}(r_{m,2})= 0.501$. Top-right panel: $\delta_1 = 0.3$ and $\delta_{2}= 0.62$, yielding to $\mathcal{C}(r_{m,1})= 0.491$, $\mathcal{C}(r_{m,2})= 0.620$. Bottom panel: $\delta_1 = 0.385$ and $\delta_{2}= 0.555$, yielding to $\mathcal{C}(r_{m,1})= 0.554$, $\mathcal{C}(r_{m,2})= 0.555$. In all cases, $q_{j}=6$ with $\delta_{c,j}=0.592$ for $j=1,2$, $t_H$ is the time of horizon crossing when the length-scale $r_{m,1}$ reenters the cosmological horizon and $\beta =2$.}
\label{fig:snapshots_C}
\end{figure}

% \subsection{Apparent horizon formation}
In this section, we study the dynamical evolution of the fluctuations in terms of the different parameters of Eq.\eqref{eq:4_dos_torres}. 
In Fig.~\ref{fig:snapshots_C}, we show the evolution of the compaction function at some specific times, for which we can differentiate three situations. 
In the top-left panel we show a case where the first peak of $\mathcal{C}$ triggers the gravitational collapse when the first peak is over-threshold $\mathcal{C}(r_{m,1})>\delta_{c,1}$ being the second one $\mathcal{C}(r_{m,2})<\delta_{c,2}$. 
In the top-right panel instead, we show a case where the second peak triggers the gravitational collapse with $\mathcal{C}(r_{m,2})>\delta_{c,2}$ being the first one under threshold $\mathcal{C}(r_{m,1})<\delta_{c,1}$. 
Notice that the time needed for the growth of the fluctuation is much larger than the case $\mathcal{C}(r_{m,2})<\delta_{c,2}$ since the size of the overlapped compaction function is much larger. In the previous cases, the peak under the threshold $\delta_{c,j}$ becomes smoothed and the mass excess disperses on the FLRW background. 
A different situation is found in the bottom panel, where both peaks are lower than $\mathcal{C}(r_{m,j})<\delta_{c,j}$. In this case, the two peaks will merge 
% during the gravitational collapse, 
and collapse even when $\mathcal{C}(r_{m,j})<\delta_{c,j}$ surpassing
the pressure gradients. Notice that the time scale is similar to the previous case since the contribution from the larger scale 
% \Blue{compaction function} 
peak
is necessary for the collapse and 
the time scale is dictated by the larger scale.

%%%%%%%%%%%%%%%%%%%

\subsection{Apparent horizon formation}

If an initial perturbation at super-horizon scales has a peak value of $\mathcal{C}$ bigger than its threshold, the perturbation will continue growing, and, at some point, a trapped surface will be formed. 
%This indicates the onset of gravitational collapse. 
To identify when trapped surfaces are formed, we have to consider the expansion $\Theta^{\pm}$ of null geodesic congruences generated by the null vector fields
$k^{\pm}$, orthogonal to a spherical surface $\Sigma$. The expansion $\Theta^{\pm}$ is defined as $\Theta^{\pm} \equiv h^{\mu \nu} \nabla_{\mu}k_{\nu}^{\pm}$, where $h^{\mu \nu}$ is the spacetime metric induced on $\Sigma$. There are two congruences: we call them inwards $k_{\mu}^{-}$ and outwards $k_{\mu}^{+}$, whose components are $k_{\mu}^{\pm}=(-A,\pm B, 0,0)$ with $k^{+} \cdot k^{-} =-2$.

If $\Theta^{\pm}<0$, the surface is called trapped, while if both are positive $\Theta^{\pm}>0$, the surface is anti-trapped. On the other hand, in the case of flat spacetime, $\Theta^{-}<0$ and $\Theta^{+}>0$, and these surfaces $\Sigma$ are called untrapped surfaces.

In our case, we have that,
\begin{equation}
\Theta^{\pm} = \frac{2}{R}(U \pm \Gamma),
\end{equation}
where $U$ is the Eulerian velocity defined as $U=\dot{R}/A$ (the dot represents time derivative $\partial / \partial t$) and $\Gamma$ is given by $\Gamma = \sqrt{1+U^{2}-2M/R}$. In spherical symmetry, we can consider that any point $(r,t)$ is a closed surface $\Sigma$ with an areal radius $R$. These points can be classified as trapped, anti-trapped and untrapped. Specifically, the transition from an untrapped to a trapped surface should satisfy $\Theta^{+}=0$ and $\Theta^{-}<0$, which corresponds to a marginally trapped surface, commonly called the ``apparent horizon'' (AH). Taking into account that $\Theta^{+}\Theta^{-} = \frac{4}{R^{2}}(U^{2}-\Gamma^{2})$, the condition for the apparent horizon is given by $U^{2}=\Gamma^{2} \Rightarrow 2M=R$.

\begin{figure}[h]
\centering
\includegraphics[width=3.0 in]{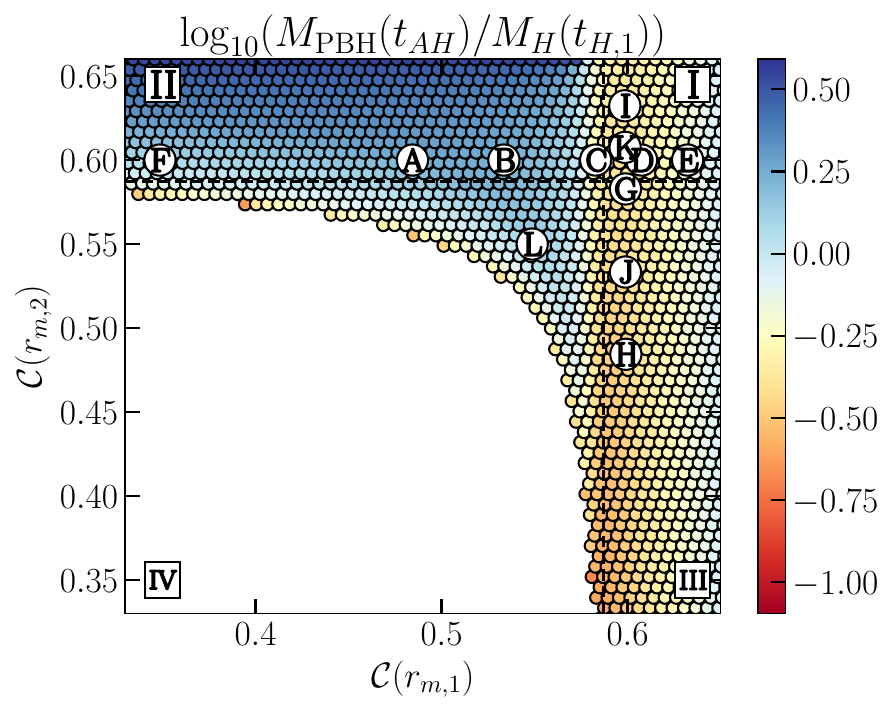}
\includegraphics[width=3.0 in]{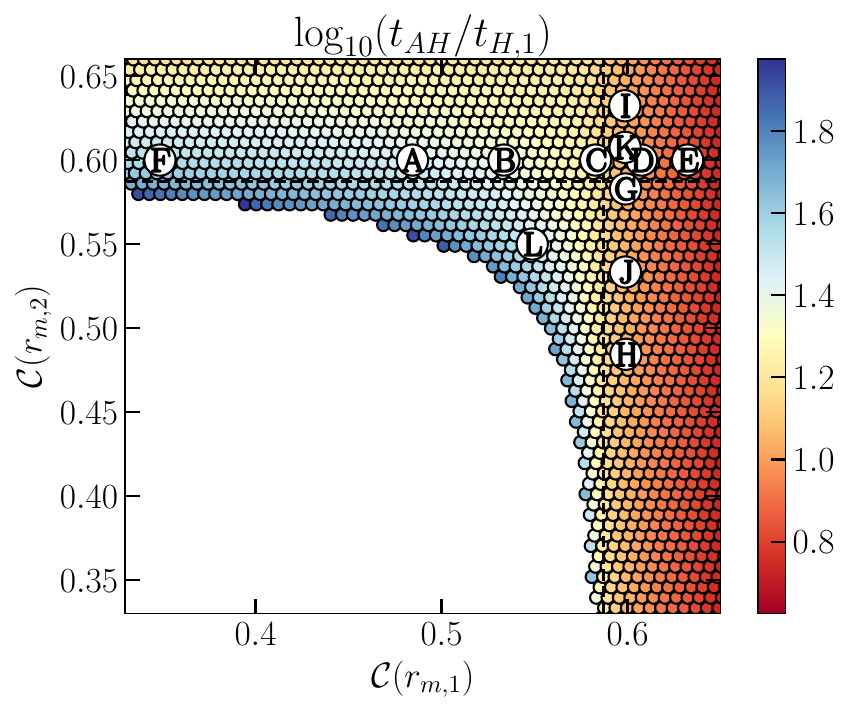}
\includegraphics[width=3.0 in]{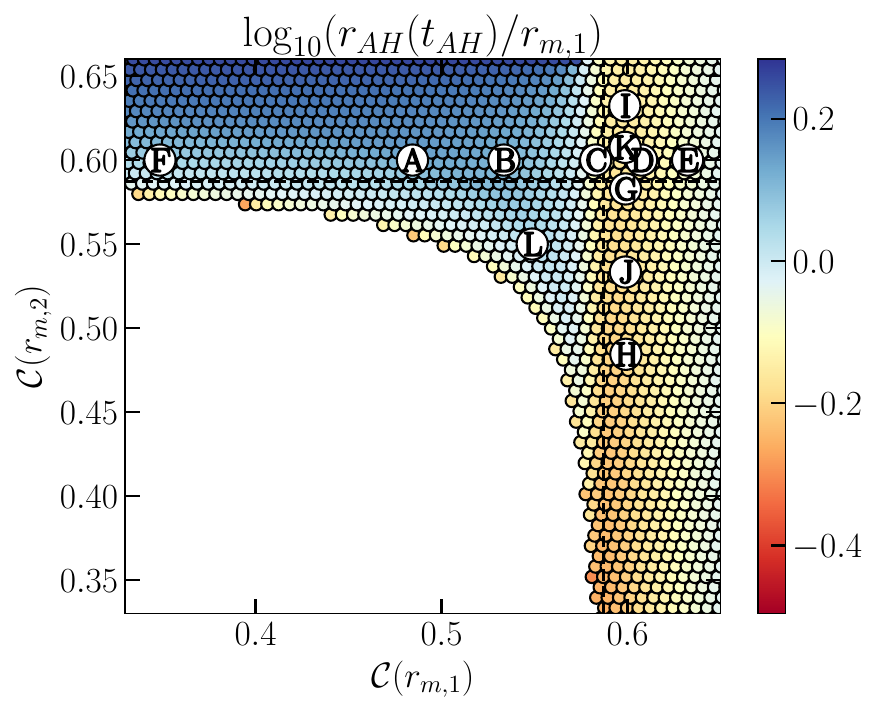}
\caption{Top-left panel: Values of the PBH mass at $t_{AH}$ in terms of the $M_{H}(t_{H,1})$ in log scale. Top-right panel: Time that takes the fluctuation to form the apparent horizon $t_{AH}$ in terms of $t_{H,1}$ and in log scale. Bottom panel: Location of the apparent horizon at its formation $t_{AH}$ in log scale. Case for $q_{1}=q_{2}=6$ and $\beta=2$. The symbols $\textrm{I}, \textrm{II}, \textrm{III}, \textrm{IV}$ in squares defines the quadrant}.
\label{fig:colour_plots_6_6_2}
\end{figure}

\begin{figure}[h]
\centering
\includegraphics[width=3.0 in]{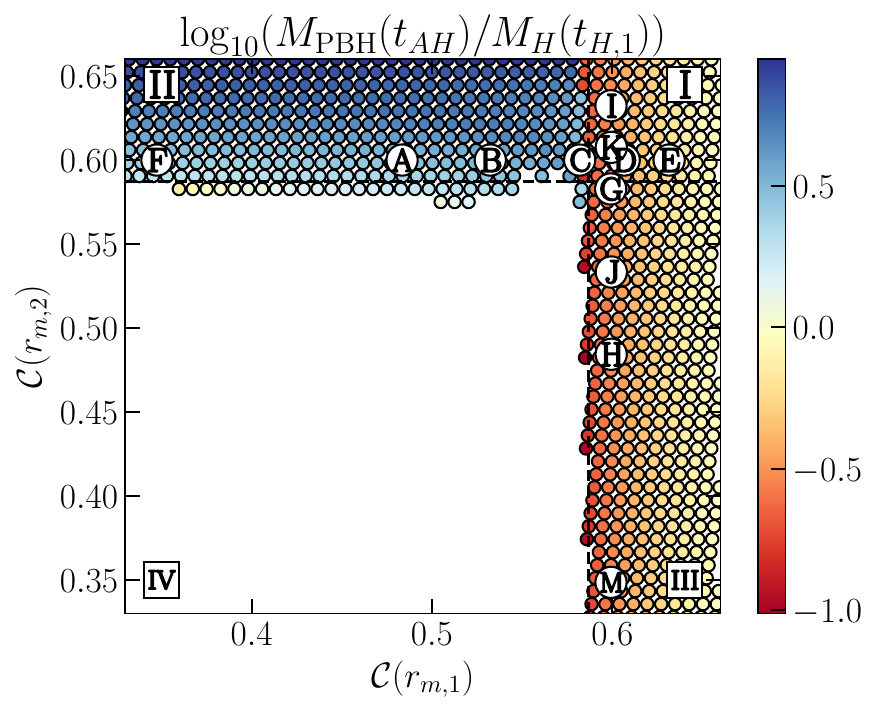}
\includegraphics[width=3.0 in]{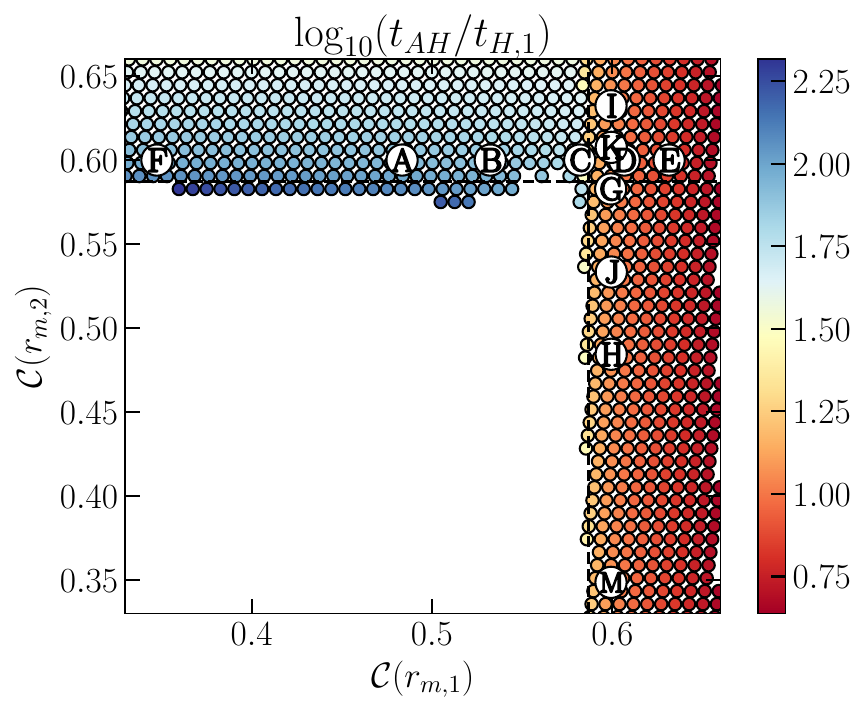}
\includegraphics[width=3.0 in]{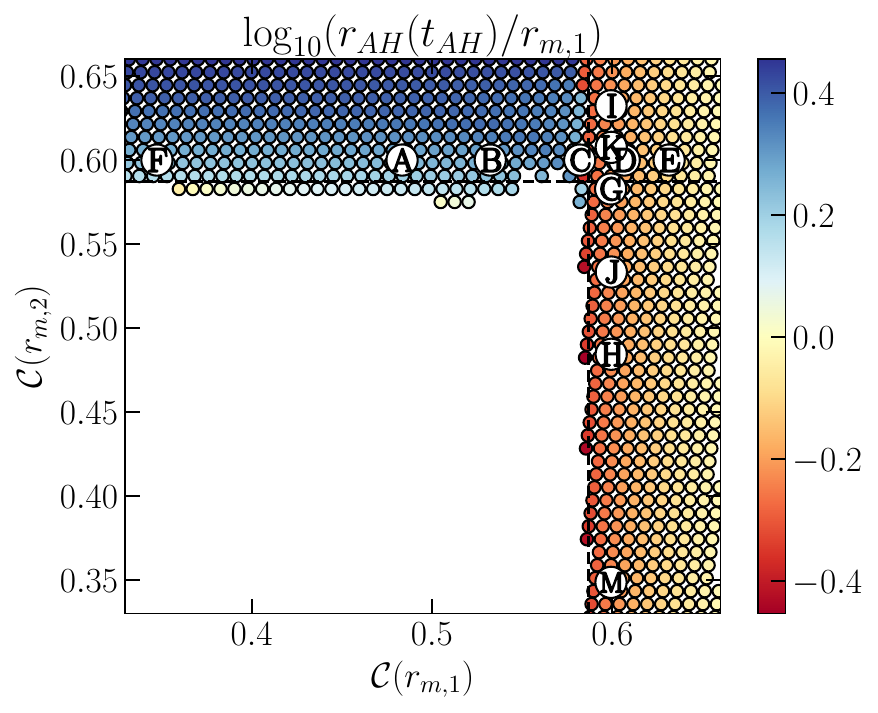}
\caption{The same caption applies from Fig.~\ref{fig:colour_plots_6_6_2}, but for the case $q_{1}=q_{2}=6$ and $\beta=3$. The symbols $\textrm{I}, \textrm{II}, \textrm{III}, \textrm{IV}$ in squares defines the quadrant.}
\label{fig:colour_plots_6_6_3}
\end{figure}

\begin{figure}[h]
\centering
\includegraphics[width=3.0 in]{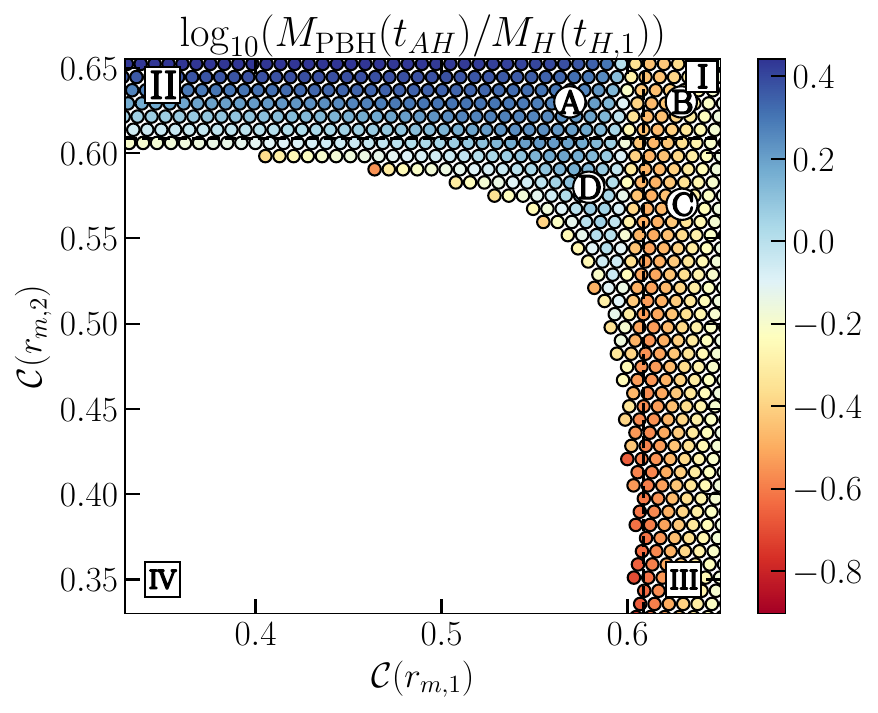}
\includegraphics[width=3.0 in]{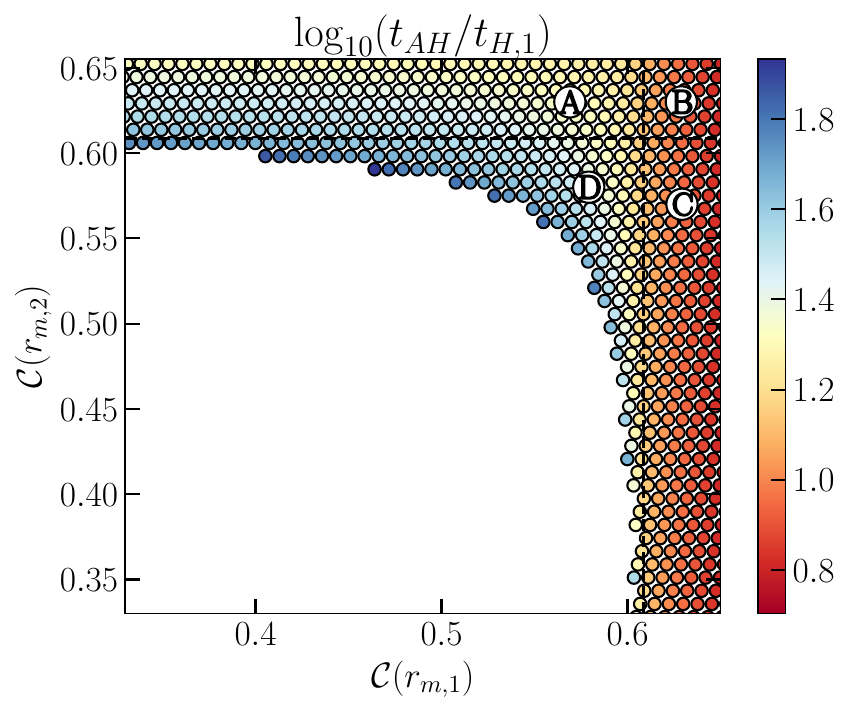}
\includegraphics[width=3.0 in]{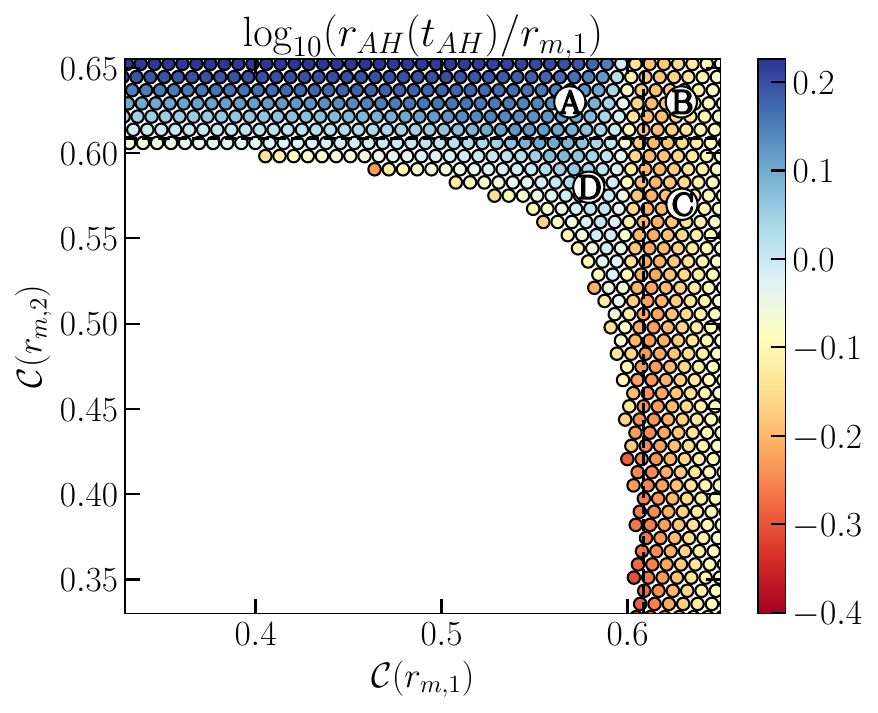}
\caption{The same caption applies from Fig.~\ref{fig:colour_plots_6_6_2}, but for the case $q_{1}=q_{2}=10$ and $\beta=2$. The symbols $\textrm{I}, \textrm{II}, \textrm{III}, \textrm{IV}$ in squares defines the quadrant.}
\label{fig:colour_plots_10_10_2}
\end{figure}

%%%%%%%%%%%%%%%
To have a better understanding of 
% how the overlapped 
% \Blue{compaction function} affects 
the dynamical evolution of the overlapped fluctuations 
% fluctuation 
until the formation of the apparent horizon, we have computed the time $t_{AH}$, the location of the apparent horizon at $t_{AH}$ namely $r_{AH}$, and the PBH mass at that time $M_{\rm PBH}(t_{AH})$. Our results, done for the different three profiles: i) $q_1=q_2=6$ and $\beta=2$, ii) $q_1=q_2=6$ and $\beta=3$ and iii) $q_1=q_2=10$ and $\beta=2$, are shown in Figs.~\ref{fig:colour_plots_6_6_2}, \ref{fig:colour_plots_6_6_3} and \ref{fig:colour_plots_10_10_2} in log-scale and with a colour mapping.

As shown in \cite{Escriva:2021pmf}, in the case of an isolated compaction function, for strong perturbations (whose amplitude is much higher than the threshold), the time of collapse is shorter, and the PBH mass is larger compared with the case of weak fluctuations (whose amplitude is close to the threshold). This behaviour is also observed in the case of having an overlapped
peaks in the
compaction function. In addition to these common tendencies, they also depend on what peak is leading the gravitational collapse since the length scale of the fluctuation plays a crucial role in the PBH mass and the time to collapse. Then we can split the phase space $(\mathcal{C}(r_{m,1}),\mathcal{C}(r_{m,2}))$ for the horizon formation in the following four regions:

\begin{itemize}
\item \textbf{Quadrant I}, $\mathcal{C}(r_{m,j})>\delta_{c,j}$ for $j=1,2$: Both peaks in $\mathcal{C}(r_{m,j})$ are over-threshold, but the one with a shorter length-scale ($r_{m,1}$) will dominate the gravitational collapse before the horizon formation, 
independently of the value of the second peak. 
We also observe that increasing the amplitude of the first peak will shorten the formation time and increase the PBH mass at the time $t_{AH}$. 
Notice that in this case, the first apparent horizon is formed in a shorter distance than $r_{m,1}$. 
As we will see later on, the secondary peak of the compaction function will play a crucial role during the accretion process from the FLRW background into the AH. 

\item \textbf{Quadrant II}, $\mathcal{C}(r_{m,1})>\delta_{c,1}$ and $\mathcal{C}(r_{m,2})<\delta_{c,2}$: Similar to the previous case, the first peak will lead the gravitational collapse whereas the second peak (under threshold) will not have a significant role for the quantities under study at $t_{AH}$. 

\item \textbf{Quadrant III}, $\mathcal{C}(r_{m,1})<\delta_{c,1}$ and $\mathcal{C}(r_{m,2})>\delta_{c,2}$: In contrast with the previous cases, now the first peak is under-threshold, and the second one will lead the gravitational collapse. Looking at the colour map, the transition from quadrant III to quadrant I is clear. 
But an effect from the first peak can be noticed
% But can be noticed an effect from the first peak 
in the boundaries of the quadrant (this is especially clear looking $M_{\rm PBH}(t_{AH})$).
The time of the apparent horizon formation 
% it takes the fluctuations to undergo the gravitational collapse 
$t_{AH}$ is larger than when the first peak leads the collapse since the length-scale is larger, implying that the PBH mass will be larger than in quadrant II. In this case, increasing the peak amplitude of the second peak will also shorten the formation time and increase the PBH mass. 

\item \textbf{Quadrant IV}, $\mathcal{C}(r_{m,1})<\delta_{c,1}$ and $\mathcal{C}(r_{m,2})<\delta_{c,2}$: Only a small region will lead to the production of PBHs. Both peaks are under the threshold $\delta_{c,j}$, and there is competition between the two peaks; indeed, the presence of a secondary peak will help the first (or vice versa) to collapse and form the apparent horizon. Interestingly, 
one can find the longest formation time $t_{AH}/t_{H}$ for this region. 
%be seen the highest formation time $t_{AH}/t_{H}$: 
Both peaks by themselves can not lead to black hole formation and it is needed that some mass excess from the second peak merges with the first peak to be able to collapse.
\end{itemize}

The qualitative behaviour in the three cases i) -- iii) is similar, but from a quantitative point of view, it depends on the specific profile considered and the ratio $\beta$. Notice that for $\beta \gg 1$, the quadrant IV will be indeed non-existent, as can be already appreciated in Fig.~\ref{fig:colour_plots_6_6_3}, since the peaks will behave as isolated as already noticed in Fig.~\ref{fig:dependence_beta2}.

\subsection{PBH mass after accretion}

The letters in the panels of Figs.~\ref{fig:colour_plots_6_6_2}, \ref{fig:colour_plots_6_6_3} and \ref{fig:colour_plots_10_10_2} are the points for what we have computed the final PBH mass, that is, the mass of the PBH at the final stage after taking into account the accretion process from the surrounding FLRW background, starting from the time $t_{AH}$. 
To obtain the temporal evolution of the PBH mass $M_{\rm PBH}(t)$, we need to use an excision technique to remove numerically the formation of the singularity that appears soon after the formation of the apparent horizon, see \cite{escriva_solo} for details. 
Then, to compute $M_{\rm PBH}$ we follow the approach already used in \cite{2017JCAP...04..050D,escriva_solo,Yoo:2021fxs}, which consists of applying the Novikov-Zeldovich accretion formula \cite{Zeldovich:1967lct} (which considers Bondi accretion \cite{1952MNRAS.112..195B}) once the accretion from the surrounding of the apparent horizon reaches a quasi-stationary regime, that is, the energy density just outside the apparent horizon decreases as in an FLRW universe. In practice, this is not satisfied at the time $t_{AH}$ when the assumption of a quasi-stationary flow is invalid. It is needed to wait for a sufficiently long time to obtain a quasi-stationary flow, and for that, we follow the criteria adopted in \cite{escriva_solo} such that $\dot{M}/(H M)\ll 1$. 

%In particular, t
The Novikov--Zeldovich mass accretion is given by, 
\begin{equation}
      \frac{dM_{\rm PBH}}{dt}  =16\pi F M_{\rm PBH}^2 \rho_{b},
    \label{eq:NZ_formula}
\end{equation}
where $F$ is a constant that corresponds to the efficiency of accretion which is commonly numerically found to be of order $\mathcal{O}(1)$ \cite{2017JCAP...04..050D,escriva_solo,Yoo:2021fxs}.

\begin{figure}[h]
\centering
\includegraphics[width=3.0 in]{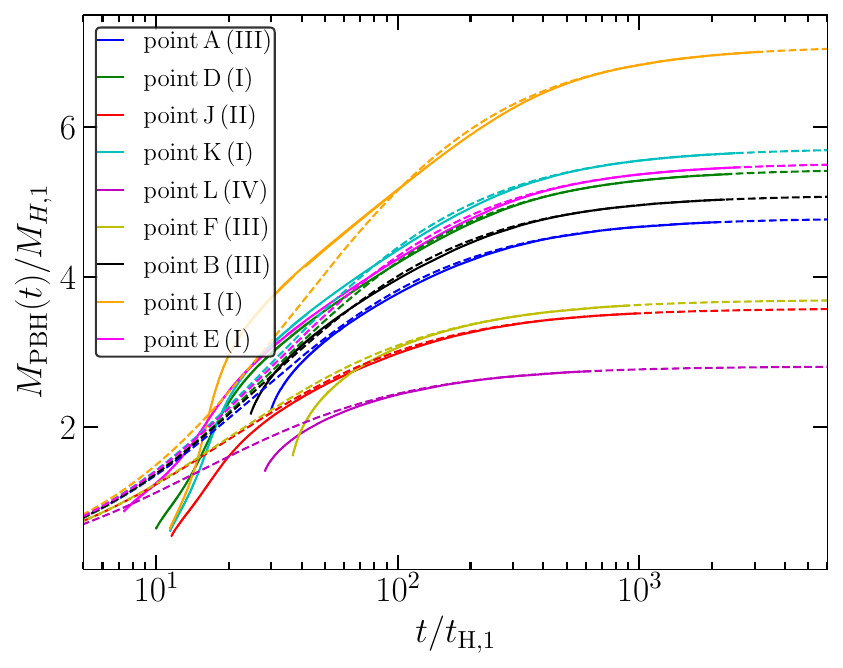}
\includegraphics[width=3.0 in]{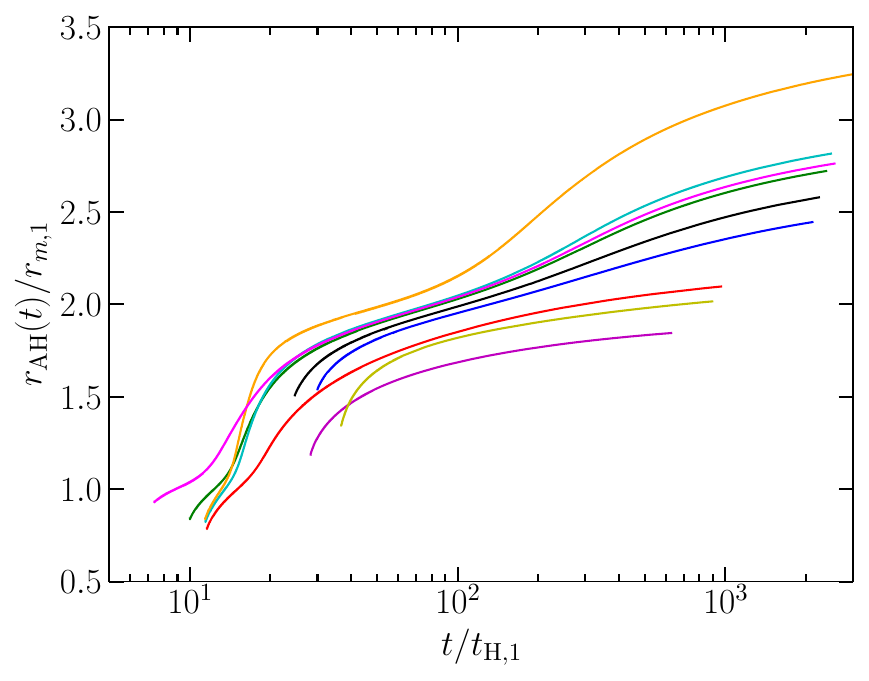}
\caption{Left-column: Time-evolution of the PBH mass for different initial conditions (solid-line) compared with the analytical fit of Eq.\eqref{eq:evolution_pbh_mas} (dashed-line). Right-column: Evolution of the location of the apparent horizon in time for different configurations. Cases from Fig.~\ref{fig:colour_plots_6_6_2}. The symbols I,II,III and IV in parentheses indicate the quadrant where the points are located.}
\label{fig:mass_evolutions}
\end{figure}

The integrated solution is,
\begin{equation}
    M_{\rm PBH}(t)=\frac{1}{\frac{1}{M_0}+\frac{3}{2}F\qty(\frac{1}{t}-\frac{1}{t_0})},
    \label{eq:evolution_pbh_mas}
\end{equation}
where $M_0$ is the PBH mass at the time $t_0$ when we consider the asymptotic approximation. Using the numerical evolution of the PBH mass $M_{\rm PBH}(t)$, we can make a fitting to obtain the parameters $M_0$ ,$t_0$ and $F$, and obtain the final PBH mass in the limit $M_{\rm PBH} = M_{\rm PBH}(t\longrightarrow \infty)$, when the increase in time of the mass will be very small. We find the values of $F$ to be $F \approx 3.5$, consistent with what was found in \cite{escriva_solo}. 
A plot of the evolution of the PBH mass for some of the points located in Figs.~\ref{fig:colour_plots_6_6_2}, \ref{fig:colour_plots_6_6_3} and \ref{fig:colour_plots_10_10_2} is shown in Fig.~\ref{fig:mass_evolutions}. 

%figuras
Let's focus on the first row of Fig.~\ref{fig:mass_evolutions}. The initial PBH mass at the time $t_{AH}$ increases substantially thanks to the accretion of the FLRW background. The dashed line corresponds to the analytical formula of Eq.\eqref{eq:evolution_pbh_mas}, which fits very well the numerical $M_{\rm PBH}(t)$ for late times. The location of the apparent horizon in time is shown in the right panel. For the points \textbf{A,B,F,L}, the second peak leads the gravitational collapse, whereas for the points \textbf{D,E,J,K}, the first peak dominates the collapse before the horizon formation. When the first peak of $\mathcal{C}$ leads the gravitational collapse, the apparent horizon will soon interact and absorb the mass excess of the second peak, which leads to an increasing monotonic behaviour of the radial coordinate $r_{\rm AH}$ as can be appreciated. 
This differs from the points \textbf{A,B,F,L} since at the formation of the AH already the mass excess from the first peak lies within $r_{AH}$, and the accretion process that follows accounts only for the tail of the secondary peak and the FLRW background. To understand the quantitative effects of the overlapping peaks in the compaction function
% \Blue{compaction functions} 
on the PBH mass, we should refer to the data shown in Table \ref{table:data_6_6_2}.

The final mass of the PBH depends on the amplitude of each peak and what peak will lead to the collapse. 
For instance, if we compare the points \textbf{I} and \textbf{E}, both have 
%a 
very similar peak values when those values are exchanged  
% its values 
$\mathcal{C}(r_{m,1}) \leftrightarrow \mathcal{C}(r_{m,2})$, but the mass is larger in the case of \textbf{I} since the secondary peak, which has a larger length scale, has a higher amplitude and dominates the gravitational collapse. 

%On the other hand, 
Even when the second peak triggers the gravitational collapse, the mass excess from the first peak also contributes to the PBH mass since the AH encloses that. For instance, for the point \textbf{C}, the final PBH mass $ \approx 5.34 M_{H}(t_{H,1})$ is larger than the respective sums of the cases having isolated peaks of the compaction function 
% \Blue{compaction functions} 
(the last two rows in Table \ref{table:data_6_6_2}) which would be $\approx 4.13 M_{H}(t_{H,1})$. In this case, the mass excess of the first peak doesn't lose its mass because of pressure gradients, compared to having the first peak isolated. Notice that for the case \textbf{I} (I quadrant), we have the case with the largest PBH mass since it has the largest $\mathcal{C}(r_{m,2})$ in comparison with other cases.

In the table, we also show the ratio $M_{\rm PBH}/M_{\rm PBH}(t_{AH})$, which gives an idea of the increase of the PBH from the time $t_{AH}$. 
When the second peak leads the gravitational collapse, the accretion increases the PBH mass by a factor $\sim 2$ (as in the case of 
an isolated peak in the compaction function), 
since the dynamical behaviour will be similar to having an isolated peak in the compaction function: only capturing the tail of the mass excess of the fluctuation that produces the apparent horizon. 
The situation is different when the first peak leads the collapse since it can capture all the mass excess from the secondary peak in the compaction function, 
as discussed before, making this ratio much larger.  

In the first row of Fig.~\ref{fig:mass_evolutions_extra} in the appendix \ref{sec:extra_figures}, 
we have the case with the same $q_1=q_2=6$ but with a larger separation between the peaks $\beta=3$ instead of $\beta=2$ (see Table \ref{table:data_6_6_3}). The qualitative behaviour is similar, but notice that since the second peak lies on much larger scales, 
the accretion 
% from the FLRW background 
is larger 
% when 
if
the first peaks 
%first 
form the AH, whereas the ratio $M_{\rm PBH}/M_{\rm PBH}(t_{AH})$ remains as a factor $\sim 2$ 
%when 
if
the second peak leads the collapse.

In the second row in Fig.~\ref{fig:mass_evolutions_extra} in the appendix \ref{sec:extra_figures}, we have shown the case $q_1=q_2=10$ with $\beta=2$ (see Table \ref{table:data_10_10_2}). The qualitative behaviour is similar to the case of the first row. However, we find differences in the PBH mass (in particular smaller in some cases) with smaller mass excess for large $q$ (comparing with equal peak amplitude $\mathcal{C}$) as noticed in \cite{Escriva:2021pmf}. It is due to large pressure gradients that reduce the effect of accretion.

\begin{table}[h]
\centering
\begin{tabular}{ccccc}
\hline
\multicolumn{1}{|c|}{$\mathcal{C}(r_{m,1})$} & \multicolumn{1}{c|}{$\mathcal{C}(r_{m,2})$} & \multicolumn{1}{c|}{$M_{\rm PBH}/M_{H}(t_{H,1})$} & \multicolumn{1}{c|}{$M_{\rm PBH}/M_{\rm PBH}(t_{AH})$} & \multicolumn{1}{c|}{Point} \\ \hline
\multicolumn{1}{|c|}{0.608} & \multicolumn{1}{c|}{0.600} & \multicolumn{1}{c|}{5.45} & \multicolumn{1}{c|}{8.36} & \multicolumn{1}{c|}{D (I)} \\ \hline
\multicolumn{1}{|c|}{0.632} & \multicolumn{1}{c|}{0.600} & \multicolumn{1}{c|}{5.53} & \multicolumn{1}{c|}{6.27} & \multicolumn{1}{c|}{E (I)} \\ \hline
\multicolumn{1}{|c|}{0.599} & \multicolumn{1}{c|}{0.632} & \multicolumn{1}{c|}{7.09} & \multicolumn{1}{c|}{10.92} & \multicolumn{1}{c|}{I (I)} \\ \hline
\multicolumn{1}{|c|}{0.599} & \multicolumn{1}{c|}{0.608} & \multicolumn{1}{c|}{5.72} & \multicolumn{1}{c|}{9.25} & \multicolumn{1}{c|}{K (I)} \\ \hline
\multicolumn{1}{|c|}{0.599} & \multicolumn{1}{c|}{0.583} & \multicolumn{1}{c|}{4.80} & \multicolumn{1}{c|}{8.09} & \multicolumn{1}{c|}{G (II)} \\ \hline
\multicolumn{1}{|c|}{0.599} & \multicolumn{1}{c|}{0.484} & \multicolumn{1}{c|}{2.82} & \multicolumn{1}{c|}{5.50} & \multicolumn{1}{c|}{H (II)} \\ \hline
\multicolumn{1}{|c|}{0.599} & \multicolumn{1}{c|}{0.533} & \multicolumn{1}{c|}{3.59} & \multicolumn{1}{c|}{6.53} & \multicolumn{1}{c|}{J (II)} \\ \hline
\multicolumn{1}{|c|}{0.484} & \multicolumn{1}{c|}{0.600} & \multicolumn{1}{c|}{4.79} & \multicolumn{1}{c|}{2.14} & \multicolumn{1}{c|}{A (III)} \\ \hline
\multicolumn{1}{|c|}{0.534} & \multicolumn{1}{c|}{0.600} & \multicolumn{1}{c|}{5.10} & \multicolumn{1}{c|}{2.34} & \multicolumn{1}{c|}{B (III)} \\ \hline
\multicolumn{1}{|c|}{0.583} & \multicolumn{1}{c|}{0.600} & \multicolumn{1}{c|}{5.34} & \multicolumn{1}{c|}{6.09} & \multicolumn{1}{c|}{C (III)} \\ \hline
\multicolumn{1}{|c|}{0.349} & \multicolumn{1}{c|}{0.600} & \multicolumn{1}{c|}{3.70} & \multicolumn{1}{c|}{2.28} & \multicolumn{1}{c|}{F (III)} \\ \hline
\multicolumn{1}{|c|}{0.549} & \multicolumn{1}{c|}{0.550} & \multicolumn{1}{c|}{2.81} & \multicolumn{1}{c|}{1.98} & \multicolumn{1}{c|}{L (IV)} \\ \hline
\multicolumn{1}{|c|}{0.599} & \multicolumn{1}{c|}{0.0} & \multicolumn{1}{c|}{0.78} & \multicolumn{1}{c|}{2.17} & \multicolumn{1}{c|}{/} \\ \hline
\multicolumn{1}{|c|}{0.0} & \multicolumn{1}{c|}{0.600} & \multicolumn{1}{c|}{3.35} & \multicolumn{1}{c|}{2.21} & \multicolumn{1}{c|}{/} \\ \hline
\multicolumn{1}{l}{} & \multicolumn{1}{l}{} & \multicolumn{1}{l}{} & \multicolumn{1}{l}{} & \multicolumn{1}{l}{} \\
\multicolumn{1}{l}{} & \multicolumn{1}{l}{} & \multicolumn{1}{l}{} & \multicolumn{1}{l}{} & \multicolumn{1}{l}{} \\
\multicolumn{1}{l}{} & \multicolumn{1}{l}{} & \multicolumn{1}{l}{} & \multicolumn{1}{l}{} & \multicolumn{1}{l}{}
\end{tabular}
\caption{Table of data for the final PBH mass estimated following Eq.\eqref{eq:evolution_pbh_mas} and corresponding to the case of Fig.~\ref{fig:colour_plots_6_6_2} for $\beta=2$ and $q_1=q_2=6$. The last two arrows correspond to the case of having isolated peaks, i.e., without having overlapped compaction functions. The symbols inside the parenthesis specify the quadrant. Notice that the numerical results are given up to some significant digits.}
\label{table:data_6_6_2}
\end{table}

\begin{table}[]
\centering
\begin{tabular}{|c|c|c|c|c|}
\hline
$\mathcal{C}(r_{m,1})$ & $\mathcal{C}(r_{m,2})$ & $M_{\rm PBH}/M_{H}(t_{H,1})$ & $M_{\rm PBH}/M_{\rm PBH}(t_{AH})$ & Point \\ \hline
0.607 & 0.600 & 10.77 & 21.36 & D (I) \\ \hline
0.632 & 0.600 & 10.88 & 13.54 & E (I) \\ \hline
0.599 & 0.632 & 16.07 & 14.40 & I (I) \\ \hline
0.599 & 0.608 & 11.77 & 27.56 & K (I) \\ \hline
0.599 & 0.583 & 8.76 & 20.83 & G (II) \\ \hline
0.599 & 0.484 & 3.11 & 7.56 & H (II) \\ \hline
0.599 & 0.533 & 5.02 & 12.02 & J (II) \\ \hline
0.599 & 0.349 & 1.44 & 3.67 & M (II) \\ \hline
0.484 & 0.600 & 9.50 & 2.29 & A (III) \\ \hline
0.532 & 0.600 & 10.24 & 2.25 & B (III) \\ \hline
0.582 & 0.600 & 10.63 & 2.81 & C (III) \\ \hline
0.348 & 0.600 & 8.17 & 2.25 & F (III) \\ \hline
\end{tabular}
\caption{Table of data for the final PBH mass estimated following Eq.\eqref{eq:evolution_pbh_mas} and corresponding to the case of Fig.~\ref{fig:colour_plots_6_6_3} for $\beta=3$ and $q_1=q_2=6$. The symbols inside the parenthesis specify the quadrant.}
\label{table:data_6_6_3}
\end{table}

\begin{table}[]
\centering
\begin{tabular}{|c|c|c|c|c|}
\hline
$\mathcal{C}(r_{m,1})$ & $\mathcal{C}(r_{m,2})$ & $M_{\rm PBH}/M_{H}(t_{H,1})$ & $M_{\rm PBH}/M_{\rm PBH}(t_{AH})$ & Point \\ \hline
0.629 & 0.630 & 5.94 & 9.36 & B (I) \\ \hline
0.629 & 0.570 & 3.87 & 6.32 & C (II)\\ \hline
0.569 & 0.630 & 5.77 & 2.45 & A (III) \\ \hline
0.579 & 0.580 & 3.40 & 2.08 & D (IV) \\ \hline
\end{tabular}
\caption{Table of data for the final PBH mass estimated following Eq.\eqref{eq:evolution_pbh_mas} and corresponding to the case of Fig.~\ref{fig:colour_plots_10_10_2} for $\beta=2$ and $q_1=q_2=10$. The symbols inside the parenthesis specify the quadrant.}
\label{table:data_10_10_2}
\end{table}

\subsection{Double apparent horizon formation}
% trapped surface
\label{sec:pbh_mass_results2}
In the previous section, we have focused on the case when the peaks in the  
compaction function are not so separated, with $\beta$ being $\mathcal{O}(1)$. 
We now consider the case when the peaks in the compaction function are highly isolated ($\beta \gg 1$) and both 
%peaks of $\mathcal{C}$ 
are over-threshold. In that case, we will 
find
% form for sufficiently late time 
% two-\Blue{apparent horizons}, 
two disconnected trapped regions in sufficiently late time
in contrast with a single 
% AH 
trapped region 
as shown in the previous cases of Fig.~\ref{fig:mass_evolutions} (see also \cite{Nakama:2014fra}). 
We call it double AH formation.

\begin{figure}[h]
\centering
\includegraphics[width=3.5 in]{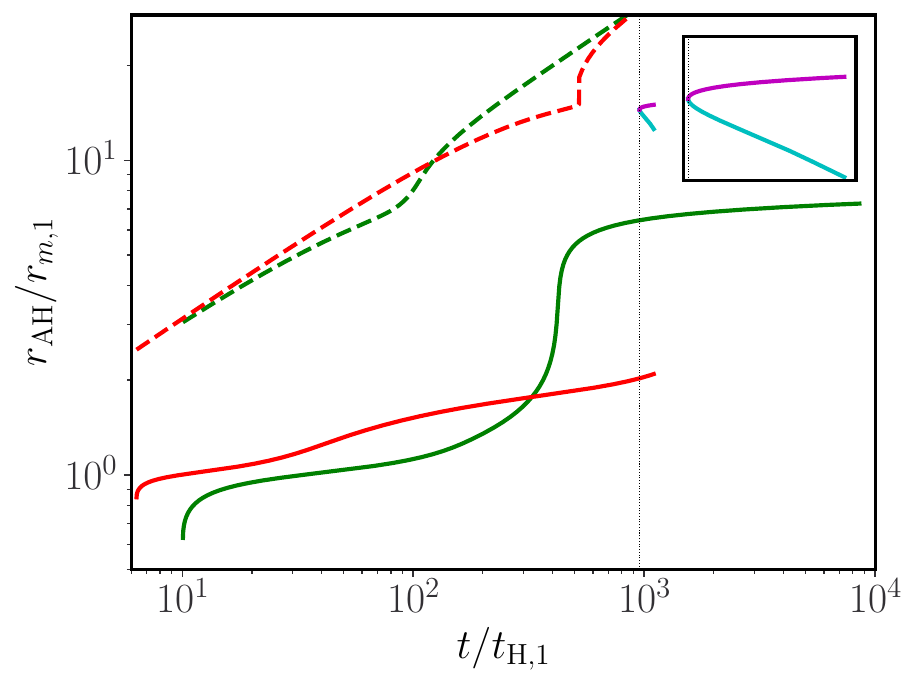}
\caption{The solid line corresponds to the location of the AH $r_{AH}$, whereas the dashed line corresponds to the cosmological horizon. The green line corresponds to the case $\beta=7$, and red for the case with $\beta=15$ (the case which corresponds to the formation of a double trapped surface). The subplot makes a zoom at the time when the second AH is formed at $t/t_{H,1} \approx 10^3$. The magenta line corresponds to the new AH, and the cyan line to the marginally trapped surface surrounded by the trapped region, which moves inwards the computational domain.}
\label{fig:evolution_horizons_very_isolated}
\end{figure}

%Specifically, first is formed the apparent horizon corresponding to the shorter fluctuation $\mathcal{C}_1$
Specifically, first the apparent horizon associated with
%corresponding to 
 % the shorter \Blue{compaction function} 
$\mathcal{C}_1$ is formed. Then, if the second peak of $\mathcal{C}$ is sufficiently separated from the first one, the AH will not be able to swallow that mass excess, allowing it to form 
%another 
a larger 
AH without a substantial interaction with the first one. In this case, a new AH is formed, which is separated from the previous one by an untrapped region $\Theta^{+}>0$ and $\Theta^{-}<0$. An example can be found in Fig.~\ref{fig:evolution_horizons_very_isolated}. 

\begin{figure}[h]
\centering
\includegraphics[width=3. in]{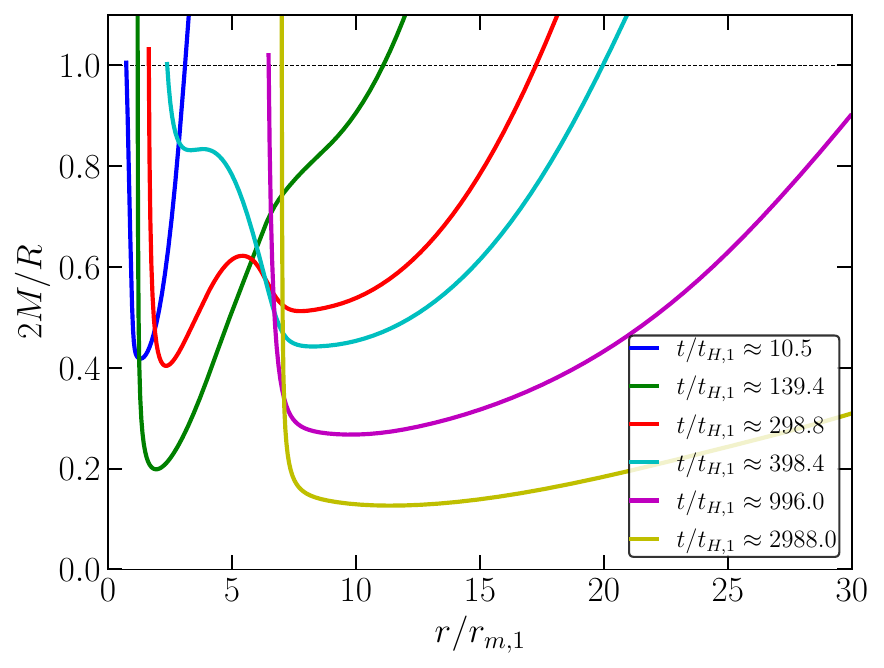}
\includegraphics[width=3. in]{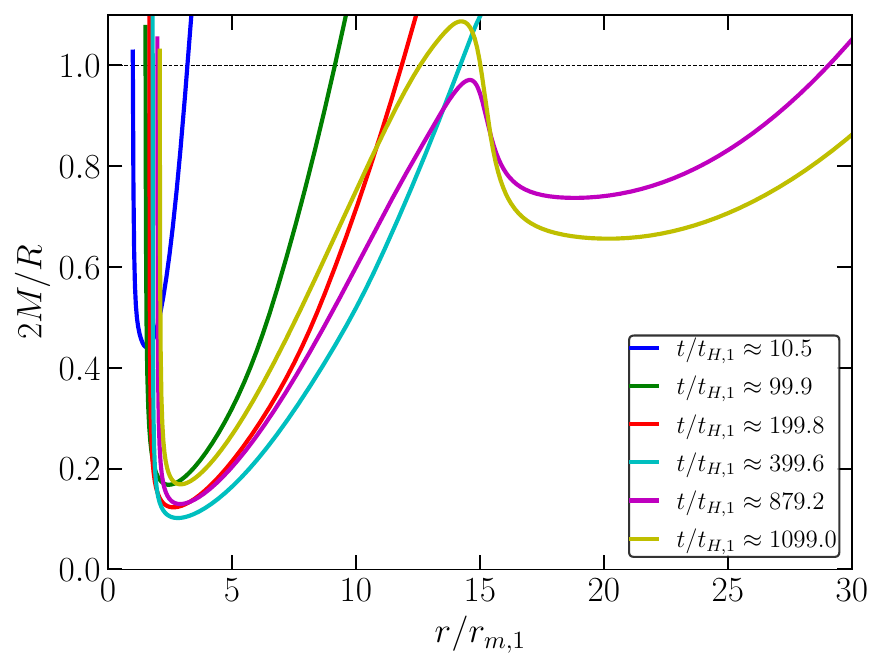}
\caption{Snapshots of the compactness $2M/R$ (solid line) at different times. Left-panel corresponds to the case with $\beta = 7$ and right-panel with the case $\beta=15$, where the double trapped surface formation is shown. See Fig.~\ref{fig:theta} in the appendix \ref{sec:extra_figures} for the corresponding snapshots of the congruences $\Theta^{\pm}$.}
\label{fig:snapshots_large_fluctuation}
\end{figure}

First, let's focus on a case similar to the previous ones in Fig.~\ref{fig:mass_evolutions} with $q_1=q_2 =6$ and $\beta=7$. 
The green lines correspond to a case with $\beta=7$, where it is clear that the AH (solid line) from the beginning captures the mass excess 
% from the overlapped \Blue{compaction function} 
associated with
$\mathcal{C}_{2}$, largely increasing its coordinate radius (see left-panels of Fig.~\ref{fig:snapshots_large_fluctuation}). In this case, there is only one connected trapped region surrounded by the marginally trapped surface (apparent horizon) 
separating the trapped region from the 
% outer 
outside 
untrapped region.
The untrapped region is also bounded by the marginally anti-trapped surface (cosmological horizon) separating the untrapped region from the outside anti-trapped region (see right-panels of Fig.~\ref{fig:snapshots_large_fluctuation}). 
% the structure of the spacetime corresponds to an anti-trapped 

The situation differs for the case $\beta=15$, for what at $t/t_{H,1} \approx 10^3$, a new AH emerges (between the magenta and cyan lines in Fig.~\ref{fig:evolution_horizons_very_isolated}) 
that surrounds the already existing one (red line). Specifically, checking the expansion of the congruences $\Theta^{\pm}$ from Fig.~\ref{fig:theta} in the appendix \ref{sec:extra_figures}, it is noticed that the magenta line in Fig.~\ref{fig:evolution_horizons_very_isolated} corresponds to the new apparent horizon, whereas the cyan line to 
the inner marginally trapped surface. 
% an inner horizon. 
Indeed, we have the same situation of the pair creation of the outer and inner horizons when the first apparent horizon is formed at $t_{AH}$ 
although it is hard to see due to the excision procedure.
To proceed with the simulation, removing the computational domain that lies inside the new apparent horizon would be necessary. 
For this situation, when $\beta \gg 1$, the PBH mass is clearly dominated by the secondary fluctuation $\mathcal{C}_{2}$. 
This result is consistent with what was found in \cite{Nakama:2014fra} for a specific curvature profile, where the phenomena called ``double PBH formation" was noticed.

%\Blue{The magenta line corresponds to the new AH, and the cyan line to the 
%marginally trapped surface surrounded by the trapped region, which moves inwards the computational domain.}

%The behaviour depends not only on the distance $\beta$, but also 
The formation of a double 
% trapped surface 
apparent horizon 
(in comparison with a single AH) depends mainly on the separation of scales between the peaks of $\mathcal{C}$ controlled by the parameter $\beta$, but also on the specific profile and the amplitudes for both peaks. Specifically, for the double apparent horizon formation, 
the first and second peaks need to have sufficient mass excess 
and the size of the first AH must be sufficiently small, so that 
the first AH does not swallow up the second peak before the second apparent horizon formation. 
It should be also noted that, in general, the realisation of the double AH formation depends on the condition of time slicing and is not gauge-independent (and thus physically observable) manner. That is, we may choose a different gauge for the time slicing in which only one connected trapped region continues to exist after the horizon formation.
\newpage
\section{Summary and Conclusions}
\label{sec:conclusions} 
In this work, we have 
%consistently 
studied the PBH formation process from the collapse of overlapping 
cosmological fluctuations.
% \Blue{compaction functions}. 
To do that, we have used numerical simulations of the gravitational collapse of the super-horizon curvature fluctuations 
considering the formalism of the compaction function $\mathcal{C}$ to set up the initial conditions.

Using a set of curvature profiles with different parameters, 
we have shown that when the length scales of two overlapped peaks in the compaction function are well separated and sufficiently decoupled, the threshold of PBH formation can be characterized by the shape around the compaction function's peak 
that is leading the gravitational collapse, where analytical formulas $\delta_c(q)$ can be accurately applied. Meanwhile, depending on the degree of overlapping between two peaks in the compaction function, the threshold can be reduced by a few percentages compared to having isolated peaks in $\mathcal{C}$. Therefore, our results confirm that the threshold for PBH formation is mainly characterized by the shape around the compaction function peak that is leading the gravitational collapse, even in the case of considering a multi-scale problem as done in this work.

Moreover, we have studied in detail the dynamics of PBH formation for different initial conditions and the effect on the PBH mass. 
When both peaks are sufficiently closer to each other, a single AH surrounded by the cosmological horizon is formed. 
When the first peak is over the threshold, the AH swallows the mass excess from the fluctuations with a larger length scale once the AH grows to such scales, and the accretion from the larger length scale fluctuation mainly gives the final PBH mass. On the other hand,
when the second peak is over the threshold (having the first one under the threshold), 
it will lead the gravitational collapse, capturing already all the mass excess inside the AH. 
The situation differs when both peaks in the compaction function are over the threshold but sufficiently decoupled. 
For that case, the second peak of $\mathcal{C}$ can have enough time to form a new AH without the interaction of the previous one forming two disconnected trapped regions. 
% a double trapped surface. 
In the context of the effects of profile dependence on PBH formation, we have shown that profile dependence can substantially affect the process of forming PBHs, regarding the threshold values, and importantly for the final PBH mass.

Future extensions of our work could be to generalize our calculations in the context of a general equation of state $w$ as in \cite{Escriva:2020tak}. Another direction can be to explore the corresponding critical regime \cite{Gundlach:2007gc} and find its dependence in terms of the different initial conditions of the double peaks in $\mathcal{C}$. Finally, using peak theory, it could be interesting to make an accurate statistical prediction of the probability of having such overlapping fluctuations. 
% \Blue{compaction function} cases. 
Although the expectation is that the probability will be very small, the reduction in the threshold values could compensate for the large reduction of the PBH production formed in this scenario. All these features are interesting future research directions. 
\begin{acknowledgments}
We thank Cristiano Germani for useful discussions and collaboration. A.E acknowledges support from the JSPS Postdoctoral Fellowships for Research in Japan (Graduate School of Sciences, Nagoya University). This work was supported in part by JSPS KAKENHI Grant Numbers JP20H05850(CY) and JP20H05853(CY). 
\end{acknowledgments}

%apendice
\appendix

\section{Details of the numerical set-up}
\label{sec:details_setup}

In this work, we have used the publicly available numerical code offered by \cite{escriva_solo,escriva_webpage} to simulate numerically the formation of PBHs from the collapse of the curvature fluctuations on the FLRW universe filled by radiation fluid ($w=1/3$). The code uses Pseudospectral methods, and we refer the reader to \cite{escriva_solo} for more details. 
Specifically, we numerically solve Misner--Sharp equations~\cite{1964PhRv..136..571M}, which describes the gravitational collapse of a perfect fluid with spherical symmetry. Solving Einstein equations taking into account Eqs.\eqref{eq:tensor-energy},\eqref{eq:2-metricsharp} and assuming a constant equation of state like $p=w \rho$ we obtain,
%There the line element is generally given by, 

\beae{\label{eq:msequations}
	\dot{U} &= -A\left[\frac{w}{1+w}\frac{\Gamma^2}{\rho}\frac{\rho'}{R'} + \frac{M}{R^{2}}+4\pi R w \rho \right], \\ 
	\dot{R} &= A U, \\ 
	\dot{\rho} &= -A \rho (1+w) \left(2\frac{U}{R}+\frac{U'}{R'}\right), \\ 
	\dot{M} &= -4\pi A w \rho U R^{2},
}
where we have used $\Gamma=R'/B$ and the lapse $A$ can be solved analytically as $A(r,t)=[\rho_b(t)/\rho(r,t)]^{1/4}$, which is smoothly connected to the FLRW background in $r\to\infty$. The initial condition on the set of Eqs.~\eqref{eq:msequations} is imposed on a super-Hubble scale so that it is connected to the perturbed metric~\eqref{eq:2-metricsharp}, as in \cite{Polnarev:2006aa,2012JCAP...09..027P,2014JCAP...01..037N}.
There, the gradient expansion method is applied to this end.
That is, the radial dependence of the Misner--Sharp equations is expanded in the gradient parameter $\epsilon(t)$ defined by
\bae{\label{eq:epsilon}
	\epsilon(t)\equiv\frac{1}{H(t)L(t)},
}
where $H(t)$ is the Hubble factor and $L(t)\coloneqq a(t)r_{m,1}$ is the length scale of the fluctuation $K_{1}$ at super-horizon scales. 
It results in the following initial conditions~\cite{Polnarev:2006aa,2012JCAP...09..027P,2014JCAP...01..037N}:
\begin{align}
\label{2_expansion}
A(r,t) &= 1+\epsilon^2(t) \tilde{A},\nonumber\\
R(r,t) &= a(t)r(1+\epsilon^2(t) \tilde{R}),\nonumber\\ 
U(r,t) &= H(t) R(r,t) (1+\epsilon^2(t) \tilde{U} ),\\ 
\rho(r,t) &= \rho_{b}(t)(1+\epsilon^2(t)\tilde{\rho}),\nonumber\\ 
M(r,t) &= \frac{4\pi}{3}\rho_{b}(t) R(r,t)^3 (1+\epsilon^2(t) \tilde{M} ),\nonumber 
\end{align}
where for $\epsilon \rightarrow 0$, we recover the (FLRW) solution. The perturbations for the tilde variables at $\mathcal{O}(\epsilon^2)$ order in gradient expansion are shown in \cite{Polnarev:2006aa,2012JCAP...09..027P,2014JCAP...01..037N}, which we summarise here:
\begin{align}
\label{eq:2_perturbations}
\tilde{\rho}&= \frac{3(1+w)}{5+3 w}\left[K(r)+\frac{r}{3}K'(r)\right] r^2_{m,1},\nonumber \\
\tilde{U} &= -\frac{1}{5+3 w} K(r) r^2_{m,1},\nonumber\\
\tilde{A} &= -\frac{w}{1+w} \tilde{\rho},\\
\tilde{M} &= -3(1+w) \tilde{U},\nonumber\\
\tilde{R} &= -\frac{w}{(1+3 w )(1+w)}\tilde{\rho}+\frac{1}{1+3 w}\tilde{U}.\nonumber 
\end{align}
Introducing the parameters $\delta_{1},\delta_{2},r_1,r_2$ we build the initial curvature profile following Eqs.\eqref{eq:compact_K},\eqref{eq:basis_C_exp} and for what we can find numerically $r_{m,1}$ (defined as the first peak in $\mathcal{C}$). Then, the initial condition can be set up.

We have ensured that for each initial configuration, the epsilon parameter $\epsilon(t_0)$ satisfies 
% is less than 
$\epsilon(t_0) \lesssim 10^{-1}$ (this ensures that the first order in gradient expansion is 
% enough 
sufficiently
accurate~\cite{2012JCAP...09..027P}), specifically we consider $\epsilon(t_0) \sim 0.1$. %Once the peak profile of the curvature perturbation is fixed as Eq.~\eqref{zeta_t} and the maximal radius $r_\um$ is found by Eq.~\eqref{eq:rm}, the initial conditions can be set up.
Note that the choice of different gauges should give equivalent numerical results up to $\calO(\epsilon^2)$ as shown in \cite{refrencia-extra-jaume}.

The initial time of our simulations is normalised as $t_0=1$ and the background conditions are given at that time by $a(t_0)=1$, $R_H(t_0)\coloneqq1/H(t_0)=2t_0$, and $\rho_{b}(t_0)=3H^2(t_0)/8\pi$.
%The characteristic time scale $t_H\coloneqq t_0(a_0r_\um/R_H(t_0))^2$ is also useful, at which time the gradient parameter reaches unity, $\epsilon(t_H)=1$.
We use two-three Chebyshev grids with size $N \approx 400$ (although for some cases the number of points is increased) to cover the different regions where higher resolution is needed. The boundary conditions are specified in \cite{escriva_solo}. The time step is chosen as $\dd{t}=\dd{t_{0}}(t/t_{0})^{1/2}$ with $\dd{t_{0}} \sim 10^{-3}$. 

Once the apparent horizon is formed, we remove part of the computational domain (excision technique) when $\mathcal{C}_{\rm max}\approx 1.1$ (the maximum value of $\mathcal{C}$) to avoid the formation of a singularity that will appear soon after $t_{AH}$. See \cite{escriva_solo} for details of the implementation, which are very similar to our case.
%In particular, it should be noted that the maximal radius 

\section{Numerical convergence}
\label{sec:num_convergence}
\begin{figure}[h]
\centering
\includegraphics[width=1.9 in]{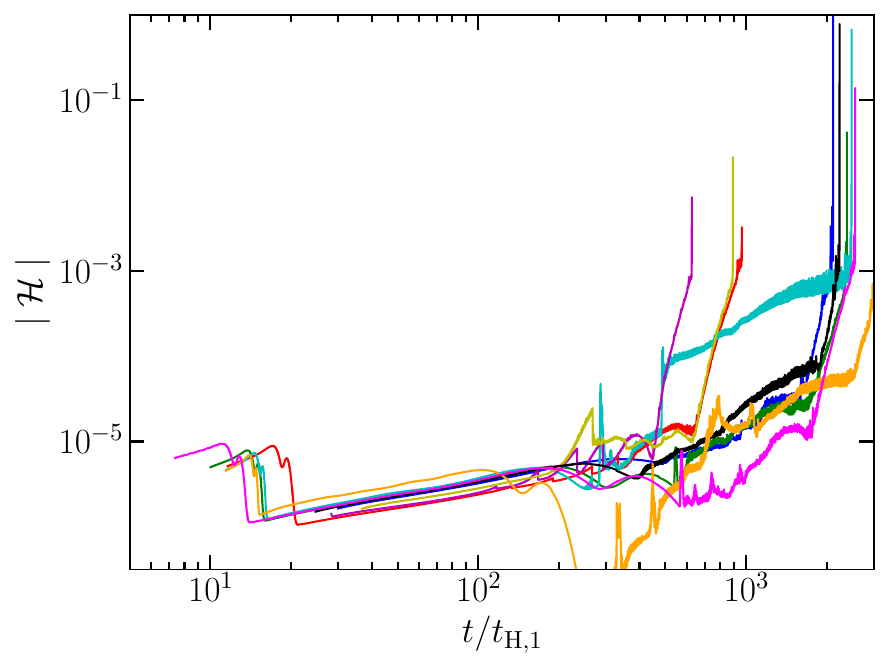}
\includegraphics[width=1.9 in]{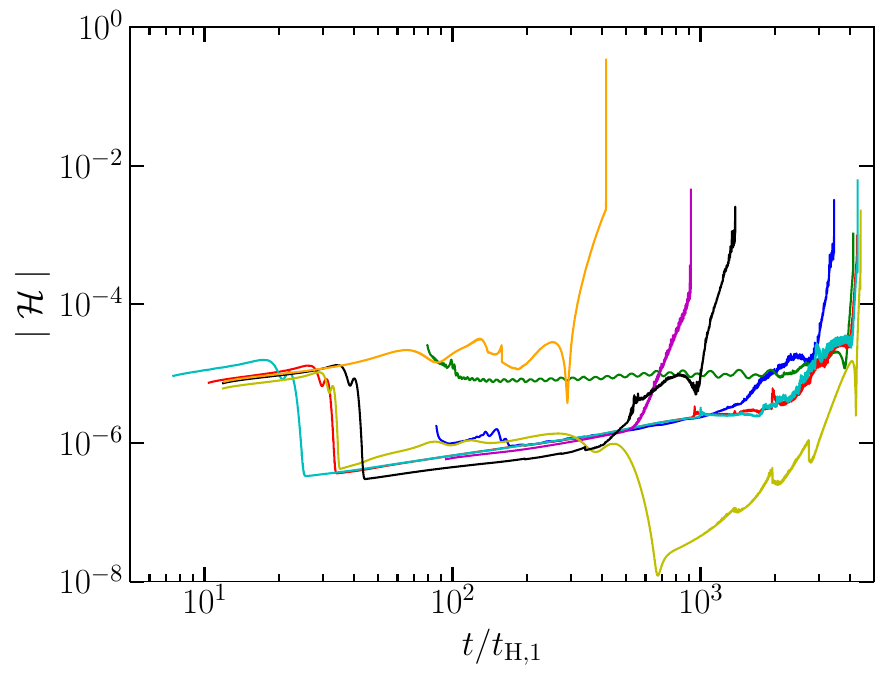}
\includegraphics[width=1.9 in]{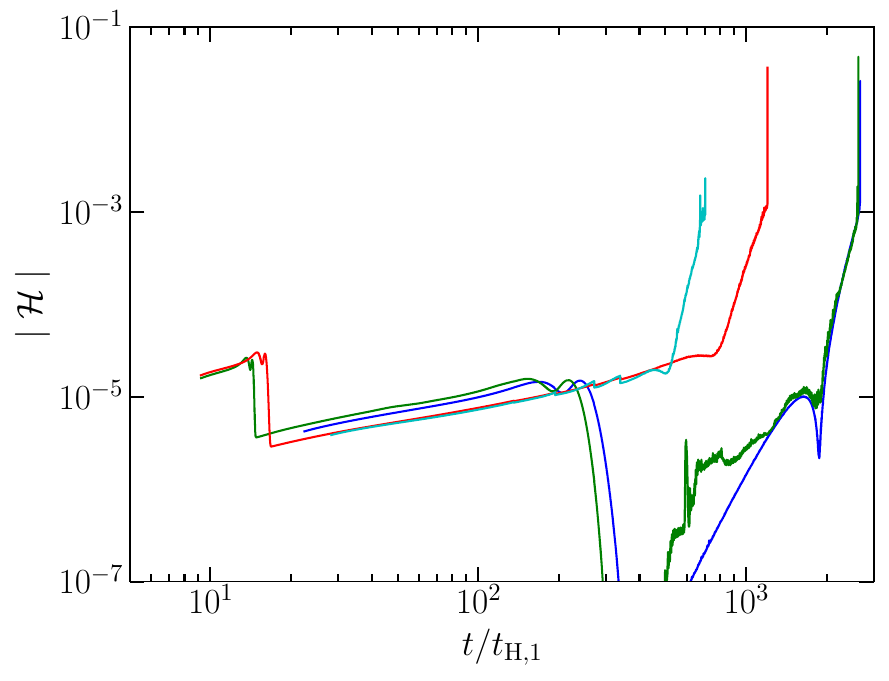}
\caption{Numerical evolution of Eq.\eqref{eq:constraint_numerical} for the cases $q_1=q_2 =6, \beta=2$ (left-panel), $q_1=q_2 =6, \beta=3$ (middle-panel) and $q_1=q_2 =10, \beta=2$ (right-panel). The evolution shown in the plot starts from the time when the apparent horizon is formed. The legend for each case is the same as in Fig.~\ref{fig:mass_evolutions}.}
\label{fig:constraints}
\end{figure}

To check the reliability of our simulations, we have performed some tests of convergence for our simulations to ensure the accuracy of the results. To do that
we use the Hamiltonian constraint equation $M' \equiv 4 \pi R' R^2 \rho$ to define the quantity,
\begin{equation}
    \mathcal{H} \equiv \frac{M'_{\rm num}-M'_{\rm def}}{M'_{\rm def}} = \frac{M'_{\rm num}/R'_{\rm num}}{4 \pi \rho_{\rm num}R^2_{\rm num}}-1,
\end{equation}
where the numerical square norm is given by,
\begin{equation}
    \mid \mathcal{H} \mid \equiv \frac{1}{N_{\rm cheb}} \sqrt{\sum_{i=1}^{N} \left(\frac{M'_i/R'_i}{ 4\pi \rho_i R^2_i}-1 \right)^2}.
    \label{eq:constraint_numerical}
\end{equation}
The $N_{\rm cheb}$ is the total number of grid points, and the sub-index $i$ refers to each grid point. For self-consistent 
% of the 
numerical simulation, $\mid \mathcal{H} \mid$ should be much smaller than one. An example of the convergence of our simulations is shown in Fig.~\ref{fig:constraints} for the estimation of the PBH mass in section \ref{sec:pbh_mass_results} (similar behaviour is found in the accuracy on determining the threshold values in section \ref{sec:threshold_results}).

The numerical fitting to the evolution of the PBH mass shown in Fig.~\ref{fig:mass_evolutions} is taken in the region where the constraints are fulfilled but at sufficiently late times where the regime of applicability of Eq.\eqref{eq:evolution_pbh_mas} is accurate. In other words, the last part of the numerical evolution where is found a substantial increment of the Hamiltonian constraints (due to an insufficient resolution of the grid \footnote{A larger numerical evolution would be possible by introducing a refined grid during the numerical evolution. However, this is unnecessary since, for this cases, the numerical evolution is sufficiently long to make an accurate fit to $M_{\rm PBH}(t)$ using Eq.\eqref{eq:evolution_pbh_mas}, which is enough for the purposes of this work.}) is not taken into account. Although that, the evolution of the PBH mass seems not to be substantially affected by that.

\newpage
\section{Supplemental figures}
\label{sec:extra_figures}
\begin{figure}[H]
\centering
\includegraphics[width=3.0 in]{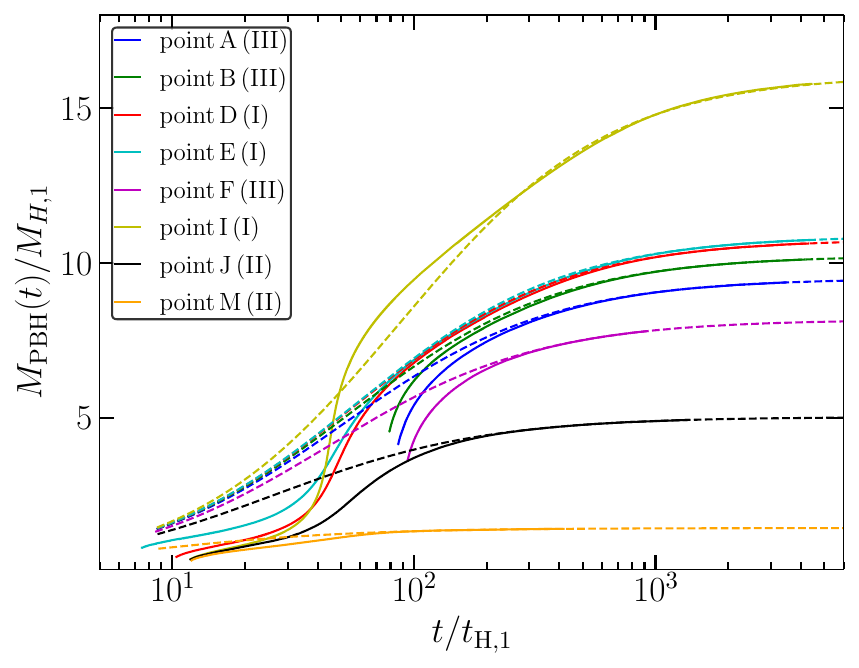}
\includegraphics[width=3.0 in]{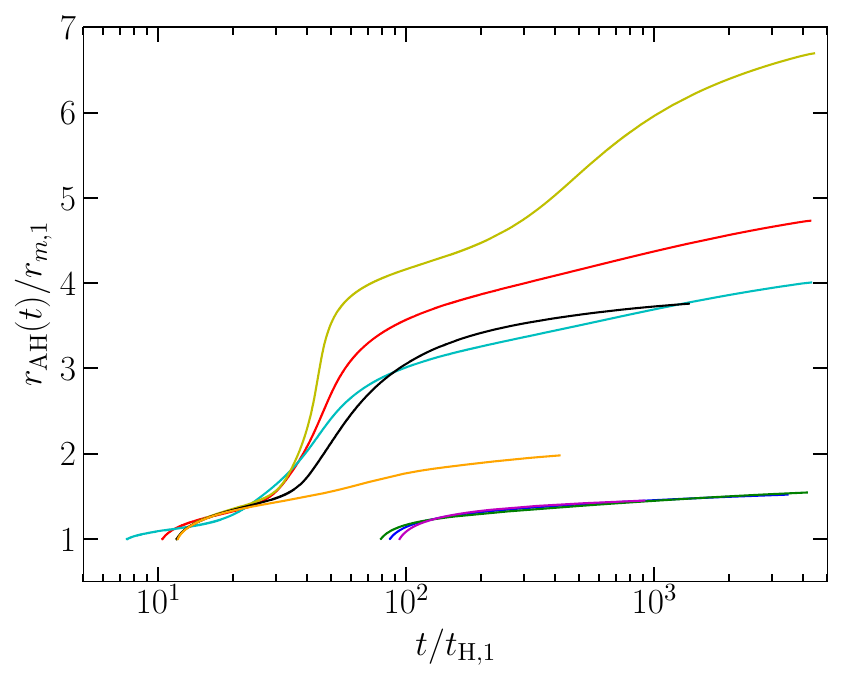}
\includegraphics[width=3.0 in]{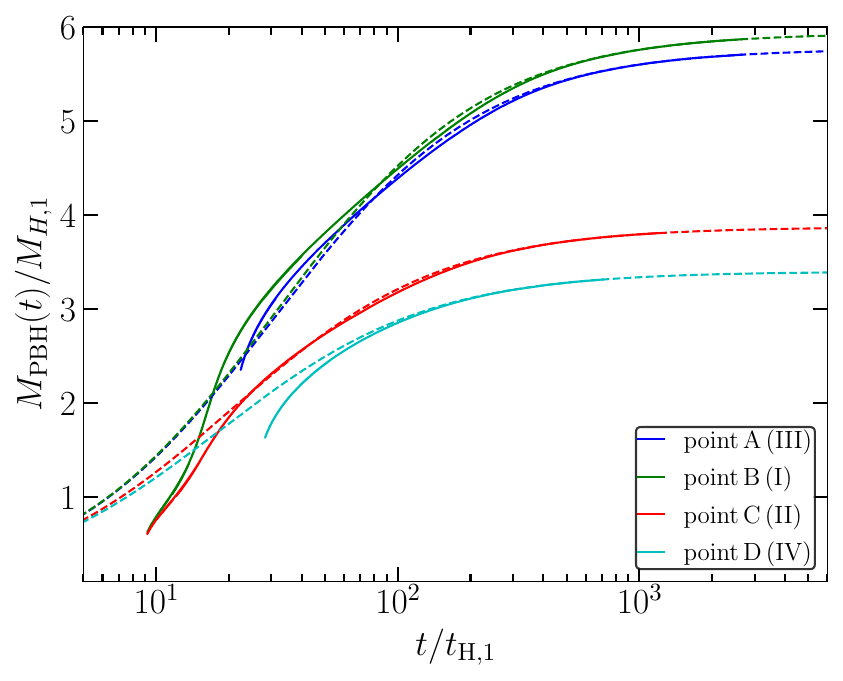}
\includegraphics[width=3.0 in]{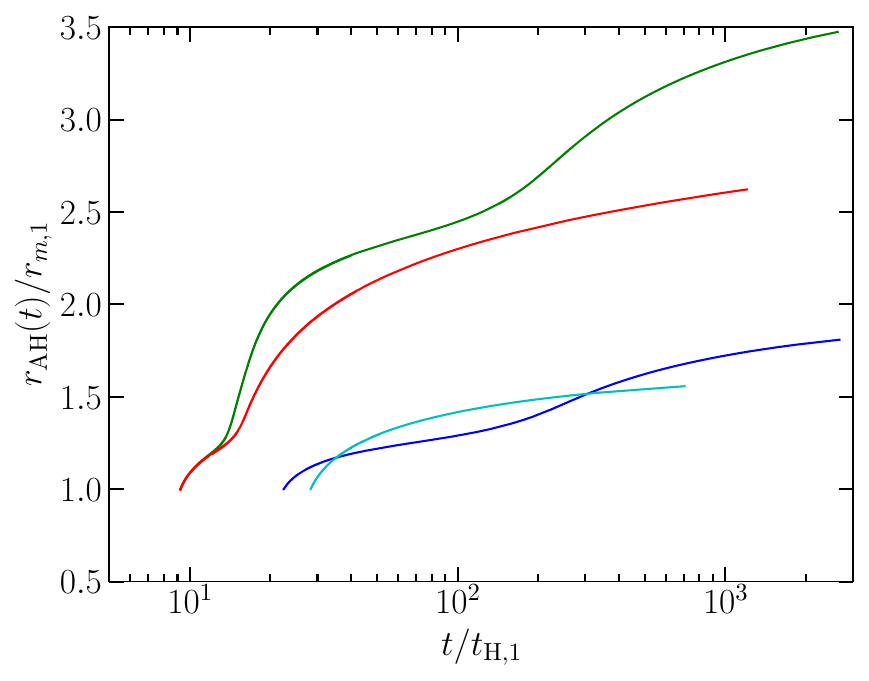}
\caption{Left-column: Time-evolution of the PBH mass for different initial conditions (solid-line) compared with the analytical fit of Eq.\eqref{eq:evolution_pbh_mas} (dashed-line). Right-column: Evolution of the location of the apparent horizon in time for different configurations. Top-panels: cases from Fig.~\ref{fig:colour_plots_6_6_3}. Bottom panel: cases from Fig.~\ref{fig:colour_plots_10_10_2}. The symbols I,II,III and IV in parentheses indicates the quadrant where the points are located.}
\label{fig:mass_evolutions_extra}
\end{figure}

\begin{figure}[H]
\centering
\includegraphics[width=3. in]{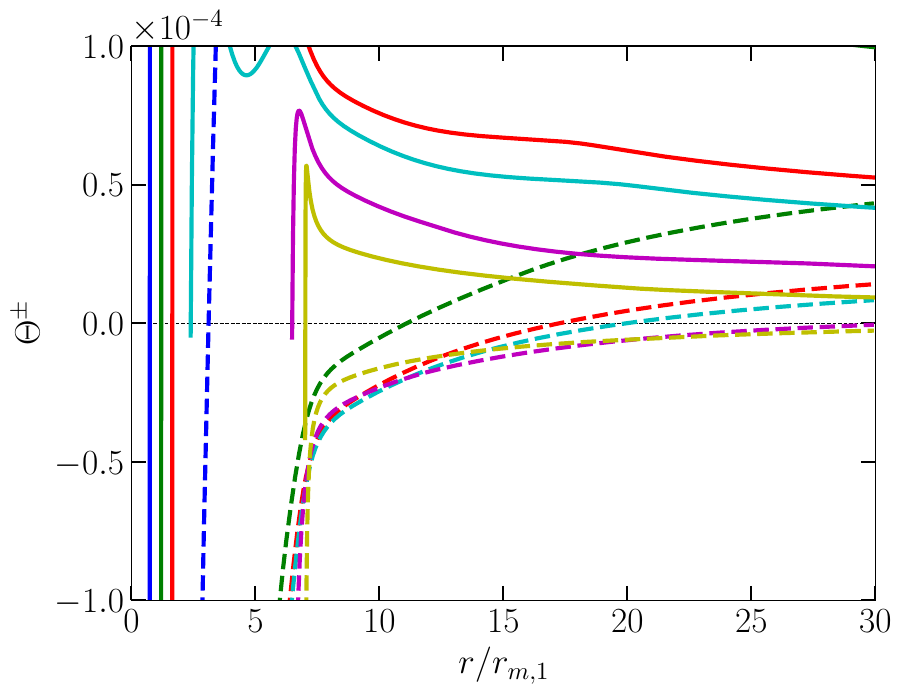}
\includegraphics[width=3. in]{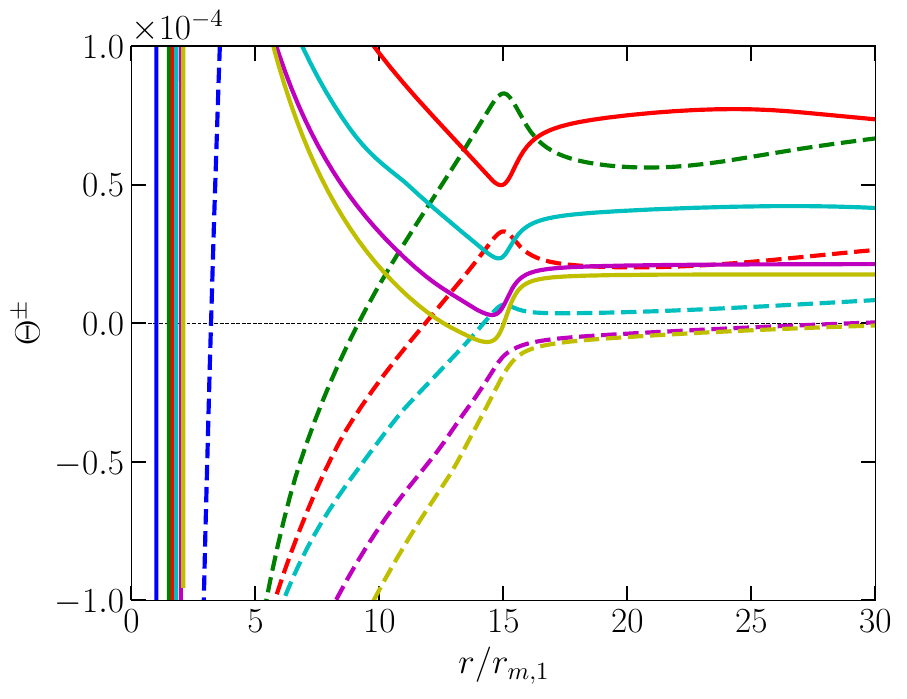}
\caption{Left panel: Snapshots of the congruences $\Theta^{+}$ (solid line) and $\Theta^{-}$ (dashed line) at different times for the case $\beta = 7$. Right panel: The same as in the left panel but with $\beta = 15$. The corresponding compactness is shown in Fig.~\ref{fig:snapshots_large_fluctuation}, with the corresponding legend which also applies to this figure.}
\label{fig:theta}
\end{figure}

In this section, we show some supplemental figures related to previous computations. In particular, in Fig.~\ref{fig:mass_evolutions_extra}, we show the time evolution of the PBH mass and the location of the apparent horizon for the cases $q_1=q_2 =6$ with $\beta=3$ (top-panels) and $q_1=q_2=10$ for $\beta=2$ (bottom-panels). In Fig.~\ref{fig:theta} we show the expansion of congruences $\Theta^{\pm}$ related with Fig.~\ref{fig:snapshots_large_fluctuation}.

\bibliographystyle{JHEP}
\bibliography{bibfile}

\end{document}